\documentclass[usenatbib]{mn2e}
\usepackage[letterpaper,margin=0.8in]{geometry}
\usepackage{natbibmnfix,fixltx2e}
\usepackage{astrojournals}
\usepackage{mathtools}
\usepackage{amsmath}
\usepackage{txfonts}
\usepackage{graphicx}
\usepackage{microtype}
\usepackage{xspace}
\usepackage{xcolor}
\usepackage[utf8]{inputenc}
\usepackage{threeparttable}
\usepackage[T1]{fontenc}
\usepackage{aecompl}

\date{}
\pagerange{\pageref{firstpage}--\pageref{lastpage}}
\pubyear{2015}

\renewcommand*{\vec}[1]{\boldsymbol{#1}}

\renewcommand{\dot}{\vec{\cdot}}
\renewcommand*{\emph}[1]{\textit{\textbf{#1}}}

\begin{document}
\label{firstpage}

\title[Conversion of IGWs]{Conversion of Internal Gravity Waves into Magnetic Waves}

\author[Lecoanet et al]{D. Lecoanet$^{1,2,3,4,5,6,7}$\thanks{E-mail: lecoanet@princeton.edu}, G. M. Vasil$^{6}$, J. Fuller$^{7,8}$, M. Cantiello$^{7}$, \& K. J. Burns$^{9}$\\
$^{1}$Physics Department, University of California, Berkeley, CA 94720, USA\\
$^{2}$Astronomy Department and Theoretical Astrophysics Center, University of California, Berkeley, CA 94720, USA\\
$^{3}$IRPHE, Marseille, 13013, France\\
$^{4}$Princeton Center for Theoretical Science, Princeton University, Princeton, NJ 08544, USA\\
$^{5}$Department of Astrophysical Sciences, Princeton University, Princeton, NJ 08544, USA\\
$^{6}$School of Mathematics \& Statistics, University of Sydney, NSW 2006, Australia\\
$^{7}$Kavli Institute for Theoretical Physics, University of California, Santa Barbara, CA 93106, USA\\
$^{8}$TAPIR, Walter Burke Institute for Theoretical Physics, Mailcode 350-17, California Institute of Technology, Pasadena, CA 91125, USA\\
$^{9}$Department of Physics, Massachusetts Institute of Technology, Cambridge, Massachusetts 02139, USA
}

\maketitle

\begin{abstract}
Asteroseismology probes the interiors of stars by studying oscillation modes at a star's surface. Although pulsation spectra are well understood for solar-like oscillators, a substantial fraction of red giant stars observed by Kepler exhibit abnormally low-amplitude dipole oscillation modes. Fuller et al. (2015) suggests this effect is produced by strong core magnetic fields that scatter dipole internal gravity waves (IGWs) into higher multipole IGWs or magnetic waves. In this paper, we study the interaction of IGWs with a magnetic field to test this mechanism. We consider two background stellar structures: one with a uniform magnetic field, and another with a magnetic field that varies both horizontally and vertically. We derive analytic solutions to the wave propagation problem and validate them with numerical simulations. In both cases, we find perfect conversion from IGWs into magnetic waves when the IGWs propagate into a region exceeding a critical magnetic field strength. Downward propagating IGWs cannot reflect into upward propagating IGWs because their vertical wavenumber never approaches zero. Instead, they are converted into upward propagating slow (Alfv\'enic) waves, and we show they will likely dissipate as they propagate back into weakly magnetized regions.  Therefore, strong internal magnetic fields can produce dipole mode suppression in red giants, and gravity modes will likely be totally absent from the pulsation spectra of sufficiently magnetized stars.
\end{abstract}

\begin{keywords}
asteroseismology; stars: magnetic field; waves; scattering
\end{keywords}

% <paper>
\section{Introduction}
\label{intro}
%%

%Asteroseismology is cool.

Many types of stars harbor strong magnetic fields that drive evolutionary processes and yield clues to prior phases of evolution. Historically, these magnetic fields have mostly been detected and studied at and above stellar photospheres \citep{Babcock_1947,Landstreet_1992,Donati_2009}, but the fields also penetrate deep into stellar interiors. Some stars may contain strong magnetic fields entirely confined within their interiors which have thus far eluded detection and engender great theoretical uncertainty. 

Asteroseismology offers the ability to probe magnetic fields deep within stars by observing stellar oscillations formed by waves that have traveled deep into the star and interacted with buried magnetic fields. Recent space-based photometry from {\it Kepler} and {\it CoRoT} has yielded exquisite asteroseismic datasets for many thousands of stars, especially red giant branch (RGB) stars. Curiously, \citet{Mosser_2011} found that roughly 20\% of RGB stars in the {\it Kepler} field exhibit low amplitude dipole oscillation modes, even though their radial and quadrupole modes appeared normal. However, \citet{Stello2016_b} later showed that low dipole mode stars also exhibit lower than normal quadrupole modes. \citet{Stello16} further showed that stars with suppressed dipole modes are preferentially higher in mass, comprised only of objects that harbored convective cores (which are capable of generating internal magnetic fields) while on the main sequence.

\citet{Fuller15} suggests low dipole mode amplitudes in RGB stars indicate strong core magnetic fields. Observable dipole oscillation modes of RGB stars have acoustic wave character in the stellar envelope, but internal gravity wave (IGW) character in the core. If a star has a strong core magnetic field (e.g., the remnant of a core dynamo), the IGW will interact with the magnetic field, partially scattering into magneto-gravity waves, which are trapped in the core of the star, explaining the depressed mode amplitudes at the surface.  Dipole modes are more suppressed than radial and quadrupole modes because more of their energy leaks into the core as IGW that can be scattered by the internal magnetic field.  They estimate the critical magnetic field required for interaction with the IGW to be
\begin{align}
\label{eqn:heuristic}
\frac{B_c}{\sqrt{4\pi \rho}}\sim \frac{1}{\sqrt{8}} \frac{\omega^2 r}{N},
\end{align}
where $\rho$ and $N$ are the density and buoyancy frequency at a radius $r$.  If the magnetic field strength exceeds $B_c$ at any point in the star, \citet{Fuller15} suggests dipole IGWs may be scattered, and thus have depressed surface amplitudes.  This paper investigates the IGW-magnetic field interaction mechanism to determine if IGWs are scattered into magneto-gravity waves when the magnetic field strength exceeds $B_c$.

Previous work on wave scattering and conversion has focused on compressible atmospheres, for application to waves near the solar surface.  In this case, one expects interaction between different wave modes when the Alfv\'en velocity $v_A$ is about equal to the sound speed $c$.  \citet{ZD81} found an exact solution to the linear, compressible, magnetized wave problem in an isothermal atmosphere with a constant magnetic field.  The solution is expressed in terms of Meijer G functions.  These can be evaluated with asymptotic methods in the limit of $v_A\ll c$ and $v_A\gg c$, i.e., before and after wave interaction.  They find IGWs entering a region of strong magnetic field almost entirely convert into slow magnetosonic waves, with the transmission coefficient proportional to $\exp(-kc/\omega)$, where $k$ and $\omega$ are the horizontal wavenumber and angular frequency of the wave.

More recently, \citet{Cally06} describes a general theory for wave conversion in the context of the WKB approximation.  As waves propagate vertically in a slowly varying background, their local vertical wavenumber changes.  Conversion can occur when two wave modes have similar vertical wavenumber.  \citet{Cally06} calculates the conversion or transmission of fast and slow magnetosonic waves from the dispersion relation, using the metaplectic formulation of \citet{Tracy93}.  This has subsequently been tested numerically \citep[e.g.,][]{McDougall07}.  However, as we describe in section~\ref{sec:heuristic}, this theory is not applicable to the interaction of IGWs with magnetic fields, as the vertical group velocity and vertical phase velocity of IGWs have opposite sign.

\citet{macgregor11} studied the interaction of IGWs with a vertically dependent, horizontal magnetic field.  As we show below, this is a singular limit of the problem of general magnetic field geometry.  They find a sufficiently strong magnetic field will reflect the IGW, but weaker fields partially reflect and partially refract IGWs.  These predictions were qualitatively confirmed by numerical simulations in \citet{rogers10}.

To test the wave scattering theory proposed in \citet{Fuller15}, we consider the evolution of IGWs propagating into a region of strong magnetic field.  In section~\ref{sec:setup}, we describe our problem setup. We solve the linearized 2D magneto-Boussinesq equations in cartesian geometry.  Section~\ref{sec:heuristic} argues that we should find perfect conversion of IGWs into slow magnetosonic modes.  This prediction is confirmed in two examples.  The first example is a background with constant magnetic field, but a vertically dependent buoyancy frequency (section~\ref{sec:constB}).  The second example is a background with a constant buoyancy frequency, but a magnetic field which is sinusoidal in the horizontal direction and exponentially decaying with height (section~\ref{sec:variableB}).  For both examples, we solve the linear problem analytically, and compare to numerical solutions.  Finally, we summarize our results and their implications for observations of mode amplitudes in section~\ref{sec:conclusions}.

\section{Problem Setup}\label{sec:setup}

The linearized magneto-Boussinesq equations \citep{Proctor82} are
\begin{align}
\rho_0\partial_t\vec{u} + \vec{\nabla} \left(p+\frac{\vec{B}\vec{\cdot}\vec{B}_0}{4\pi}\right) & = -g\rho \vec{e}_z + \frac{1}{4\pi} \left(\vec{B}_0\vec{\cdot}\vec{\nabla}\vec{B}+\vec{B}\vec{\cdot}\vec{\nabla}\vec{B}_0\right), \label{eqn:momentum} \\
\vec{\nabla}\vec{\cdot}\vec{u} & = 0, \\
\partial_t\rho & = \frac{\rho_0N_0^2}{g}\vec{e}_z\vec{\cdot}\vec{u}, \\
\partial_t\vec{B} & = \vec{B}_0\vec{\cdot}\vec{\nabla}\vec{u} - \vec{u}\vec{\cdot}\vec{\nabla}\vec{B}_0, \label{eqn:induction}
\end{align}
where $\vec{u}$ and $\vec{B}$ are the Eulerian fluid velocity and magnetic field perturbations, $p$ and $\rho$ are the Eulerian pressure and density perturbations, $g$ is the strength of gravity, and $\vec{e}_z$ is the unit vector in the (vertical) direction of gravity.  We define $z$ to be height (rather than depth), such that large $z$ corresponds to larger radii.

The background state is described by a background magnetic field $\vec{B}_0$, the background density is $\rho_0+\overline{\rho},$ where $\rho_0$ is constant, and $\overline{\rho}\ll\rho_0$ gives the background stratification via the buoyancy frequency,
\begin{align}
N_0^2 = - g \frac{\partial_z \overline{\rho}}{\rho_0}.
\end{align}
For the background to be in equilibrium, we require $(\vec{\nabla}\vec{\times}\vec{B}_0)\vec{\times}\vec{B}_0=0$ and $\vec{\nabla}\vec{\cdot}\vec{B}_0=0$.  The background magnetic fields we consider in this paper satisfy the condition $\vec{\nabla}\vec{\times}\vec{B}_0=0$.

We normalize the lengths and timescales in our problem by setting the vertical extent of the domain and the frequency of the forced IGW to unity.  Furthermore, we use the normalization $\sqrt{4\pi\rho_0}=1$ so magnetic fields have units of velocity.  We quote magnetic field strengths using these dimensionless units.  However, in all cases, the magnetic field amplitude is similar to the critical amplitude, which can be about $10^5 \ {\rm G}$ for typical RGB stars \citep{Fuller15,Cantiello_2016}.

In this work, we neglect the effect of rotation.  In the absence of magnetic fields, rotation turns internal gravity waves (IGWs) into mixed internal inertial-gravity waves.  When magnetic fields are introduced, Alfv\'enic waves become ``magnetostrophic'' waves (in analogy to geostrophic motions). The inertial-gravity waves become magneto-Poincar\'e waves, or magneto-Rossby waves if global curvature effects are included \citep[e.g.,][ and references within]{Mathis11}.

We solve these equations in two cartesian dimensions labeled $x$ and $z$.  We use $u$ and $w$ to denote the horizontal and vertical velocity, respectively.  Although we are interested in dipole IGWs which are global oscillation modes, we restrict our attention to cartesian geometry for its simplicity.  This allows us to derive analytic solutions to the linear wave problem.  We discuss the possible effects of three dimensionality in section~\ref{sec:conclusions}.

The Boussinesq approximation assumes the typical vertical lengthscale of fluid motions is much smaller than a pressure scaleheight.  For IGWs, this means the vertical wavelength of the waves must be smaller than a pressure scaleheight.  Low frequency (relative to $N_0$) IGWs have a vertical wavelength smaller than the horizontal wavelength (comparable to a pressure scaleheight) by $\omega/N_0$.  In typical RBG stars, waves of interest have $\omega/N_0\sim 10^{-2}$ at the H-burning shell.  However, at the top of the He core, $\omega/N_0\sim 1$, so the Boussinesq approximation will not be valid.

To confirm our analytic solutions, we also simulate equations~\ref{eqn:momentum}--\ref{eqn:induction} using the Dedalus\footnote{Dedalus is available at http://dedalus-project.org.} pseudo-spectral code \citep{Burns17}.  To excite IGWs near the top of our domain, we add a forcing term to the density equation
\begin{align}
F=\sin(\omega t - k_h x)\exp\left(-\frac{(z-z_0)^2}{\Delta z^2}\right),
\end{align}
where $\omega$ and $k_h$ are the frequency and horizontal wave number of the forced wave, $z_0$ is the forcing height, and $\Delta z$ is the width of the forcing.  To prevent reflections, we also add damping layers to the top and bottom of the domain, where we damp all perturbations to zero using Newtonian relaxation.  Although we use this as a numerical trick, physically, it would correspond to the effects of a porous medium on either end of the domain.  We parameterize this effect with the damping rate
\begin{align}
D_N(z) = \frac{1}{2\tau} \left[ \tanh\left(\frac{z-z_{\rm top}}{\Delta z}\right) + \tanh\left(\frac{z_{\rm bot}-z}{\Delta z}\right) +2 \right],
\end{align}
where $z_{\rm top}$ and $z_{\rm bot}$ are the heights of the top and bottom damping layers, and $\tau$ is the damping time.  This strongly damps perturbations on timescales longer than $\tau$ above $z_{\rm top}$ (and below $z_{\rm bot}$), with very little damping in between.  The damping enters our equations as shown in appendix~\ref{sec:dedalusequations}.

The equations are solved on a domain spanning $(0,L_x)$ in the $x$ direction and $(0, L_z)$ in the $z$ direction.  We run with a resolution of $(N_x,N_z)$ spectral modes with 3/2 dealiasing.  For the constant $\vec{B}_0$ problem we use Fourier modes in the $x$ direction and Chebyshev modes in the $z$ direction.  For the variable $\vec{B}_0$ problem, we use Fourier modes in both $x$ and $z$ directions.  For timestepping, we use a two stage, second order implicit-explicit Runge-Kutta method \citep{Ascher97} with a uniform timestep.  Simulation parameters are reported in table~\ref{tab:sims}.

\begin{table*}
  \caption{Parameters for our simulations with constant $\vec{B}_0$ but variable $N_0$, and constant $N_0$ but variable $\vec{B}_0$.  The size of the domain is $L_x$ by $L_z$, and we use a resolution of $N_x$ ($N_z$) modes in the horizontal (vertical) direction.  The waves are forced at $z_0$ with frequency $\omega=1$ and horizontal wavenumber $k_h$.  The background buoyancy frequency profile is given by $N_0$.  The dimensionless strength of the background magnetic field is $B_0$.  The size of the forcing region is $\Delta z$.  We also including damping layers below $z_{\rm bot}$ and above $z_{\rm top}$, with a damping timescale $\tau$.  We use a two stage, second order implicit-explicit Runge-Kutta timestepping method \citep{Ascher97} with timestep $\Delta t$.}\label{tab:sims}
{\centering
  \begin{tabular}{cccccccccccc}
    \hline
    Simulation & $(L_x,L_z)$ & $(N_x,N_z)$ & $z_0$ & $N_0$ & $B_0$ & $k_h$ & $\Delta z$ & $z_{\rm bot}$ & $z_{\rm top}$ & $\tau$ & $\Delta t$ \\
    \hline
    \hline
    Variable $N_0$ & $(0.25, 1)$ & $(16, 512)$ & 0.875 & 2(5-4z) & $5.25\times10^{-3}$ & $16\pi$ & 0.005 & 0.075 & 0.925 & 3 & 0.01 \\
    Variable $\vec{B}_0$ & $(6, 1)$ & $(256, 2048)$ & 0.85 & 40 & $5.5\times10^{-4}$ & $4\pi/6$ & 0.025 & 0.075 & 0.925 & 1 & 0.00825
  \end{tabular}
}
\end{table*}

The first problem (section~\ref{sec:constB}) has a constant background magnetic field and a linear background buoyancy frequency profile (figure~\ref{fig:background_constB}).  The implementation of this problem in Dedalus is described in appendix~\ref{sec:dedalusequations}.  The simulation parameters are listed in table~\ref{tab:sims} in the row labelled ``Variable $N_0$.''  We use a buoyancy frequency profile $N_0=2(5-4z)$ and a magnetic field strength of $B_0=5.25\times10^{-3}$.

In the second problem (section~\ref{sec:variableB}), we assume a background magnetic field of the form
\begin{align}
\label{eqn:B}
\vec{B}_0 = B_0 \sin (k_B x)\exp(-k_B z)\vec{e}_x + B_0 \cos(k_B x)\exp(-k_B z)\vec{e}_z,
\end{align}
where $k_B$ is the wavenumber of the horizontal oscillations of the magnetic field. This field satisfies $\vec{\nabla}\vec{\times}\vec{B}_0=0$ and $\vec{\nabla}\vec{\cdot}\vec{B}_0=0$.  Because we expand the problem as a Fourier series in the $z$ direction, we must have a periodic background magnetic field.  Thus, we taper the background magnetic field to zero in the damping regions by multiplying $\vec{B}_0$ by $0.5[\tanh(z-0.025/0.00625)+\tanh(0.975-z/0.00625)]$.  To simplify the analysis, we assume $N_0$ is a constant.  The implementation of this problem in Dedalus is described in appendix~\ref{sec:dedalusequations}.  The simulation parameters are listed in table~\ref{tab:sims} in the row labelled ``Variable $\vec{B}_0$.''  We use a buoyancy frequency $N_0=40$ and a magnetic field strength of $B_0=5.5\times10^{-4}$ with wavenumber $k_B=2\pi/6$.

The variable magnetic field problem is much more computationally difficult because small scale features develop which require high resolution.  The system drives energy to the smallest possible scales.  In order to run a resolved simulation, we require some dissipation to regularize the small scales.  To dissipate energy at small scales, we zero out the amplitudes of all modes with horizontal wavenumber $|k_h|>k_{h,{\rm max}}/2$ or with vertical wavenumber $|k_z|>k_{z,{\rm max}}/2$ at every timestep.  $k_{h,{\rm max}}$ and $k_{z,{\rm max}}$ are the maximum horizontal and vertical wavenumbers in the simulation.  This guarantees our simulations are well resolved without introducing any damping on larger scales.

\section{Heuristic Solution}\label{sec:heuristic}

\begin{figure}
\includegraphics[width=\columnwidth]{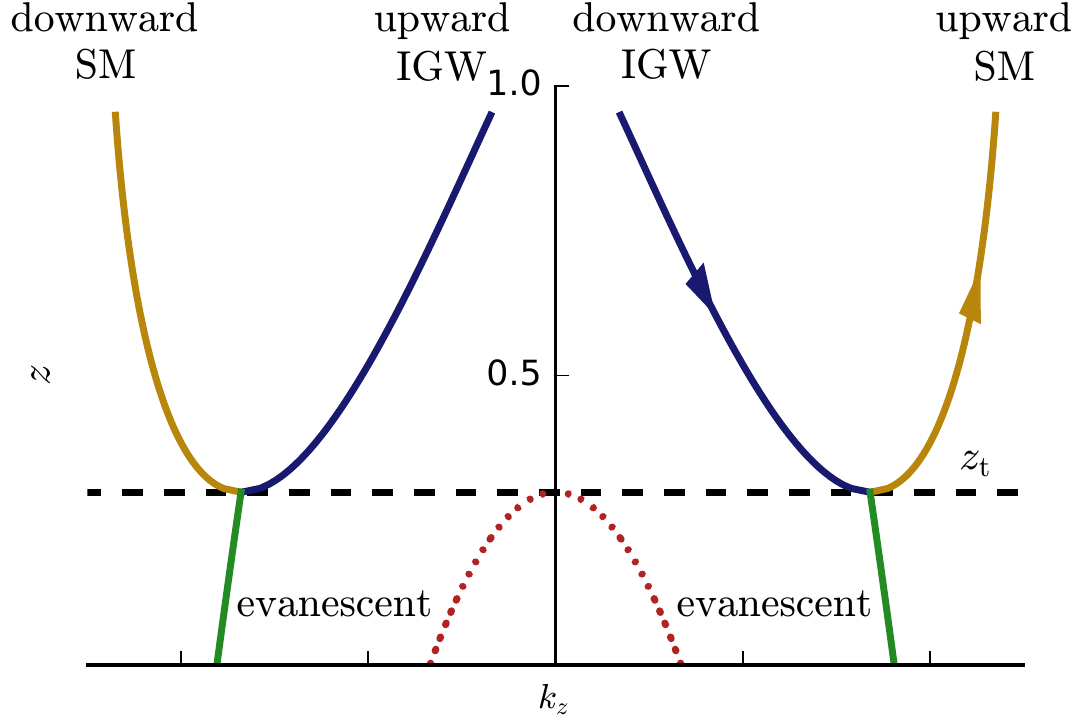}
\caption{\label{fig:schematic} The local vertical wavenumber $k_z=\theta'$ in the WKB approximation for the variable $N_0$ problem in section~\ref{sec:constB}.  At large $z$, there are four wave modes: two IGWs (blue lines), and two slow magnetosonic (SM) waves (yellow lines).  A downward propagating IGW has positive $k_z$ because the vertical group velocity and vertical wavenumber have opposite sign.  There are no propagating waves below the turning point $z_{\rm t}$ (dashed line)---the green (red) lines show the real (imaginary) part of $k_z$.  When a downward propagating IGW approaches $z_{\rm t}$, it can only convert into an upward propagating SM wave.  There is no reflection into upward propagating IGWs because the $k_z$'s of the two IGWs are not equal to each other at $z_{\rm t}$.}
\end{figure}

Although dipole IGWs are large-scale in the horizontal directions, since $\omega/N$ is small, their vertical variation is very rapid.  This suggests the Wentzel--Kramers--Brillouin (WKB) approximation\footnote{This approximations goes by many names including: the phase-integral approximation, the Carlini approximation, the Liouville--Green approximation, the Rayleigh--Gans--Jeffreys approximation, etc. \citep[e.g.,][]{dingle73}.  See \citet{Gough07} for a review of its application to stellar oscillations.} is an effective description for the waves in the vertical direction (but not the horizontal directions).  We assume the wave fluctuations (e.g., vertical velocity) can be written as
\begin{align}
w(x,z) = A(x,z)\exp\left[i\theta(z) + i k_h x - i \omega t \right],
\end{align}
where $A(x,z)$ is a slowly varying amplitude, and $\theta$ is the rapidly varying phase of the wave.  The local vertical wavenumber is $\partial_z\theta=\theta'$.  We assume the horizontal wavenumber $k_h$ and frequency $\omega$ are positive.

At large radii in the star, where the magnetic field is weak, there are four linear wave modes.  The two IGWs have local vertical wavenumbers
\begin{align}
\label{eqn:IGWdisp}
k_z \sim \mp\frac{N_0 k_h}{\omega},
\end{align}
where we have assumed $|k_z|\gg |k_h|$.  A wave's energy propagates along its group velocity, whereas the phase velocity is along $\vec{k}$. Crucially, the {\it upward} propagating IGW has {\it negative} $k_z$, and the {\it downward} propagating IGW has {\it positive} $k_z$.  This is because the group and phase velocities of IGWs are perpendicular, so the vertical wavenumber and vertical group velocity have opposite sign.\footnote{The only dimensionful quantity associated with IGWs is the buoyancy frequency $N_0$.  Thus, the dispersion relation must take the form $\omega=f(N_0)g(\vec{\hat{k}})$, where $f$ and $g$ are functions, and $\vec{\hat{k}}=\vec{k}/|\vec{k}|$.  The group and phase velocity are perpendicular because $\vec{k}\dot(\partial\vec{\hat{k}}/\partial\vec{k})=\sum_i k_i(\partial\vec{\hat{k}}/\partial k_i)=0$.}  This misalignment is the reason IGWs completely convert into slow magnetosonic (SM) waves.  The other two modes are SM waves with local vertical wavenumbers
\begin{align}
\label{eqn:SMdisp}
k_z \sim \pm \frac{\omega\sqrt{4\pi \rho_0}}{B_{0z,{\rm rms}}},
\end{align}
where $B_{0z,{\rm rms}}$ is the horizontal root-mean-square of the vertical background magnetic field.  We assume $k_z\gg k_h$ so that the horizontal component of the background magnetic field is unimportant.  The upward (downward) propagating SM wave has positive (negative) $k_z$.

If $B_{0z,{\rm rms}}$ and $N_0$ are small, then the $k_z$'s of the IGWs will be smaller in magnitude than the $k_z$'s of the SM waves.  However, as the wave propagates downward into the star, both $B_{0z,{\rm rms}}$ and $N_0$ increase.  This causes the IGW's $k_z$ to increase and the SM wave's $k_z$ to decrease.  When the two $k_z$'s approach each other, the two wave modes begin to interact.  This behavior is depicted in figure~\ref{fig:schematic}.  The two sets of modes have equal $k_z$ at the turning point.  Below the turning point, the modes' vertical wavenumbers are complex (but not purely imaginary), and thus are evanescent.

We are interested in three wave modes above the turning point.  There is the incident, downward propagating IGW, the reflected upward propagating IGW, and the converted upward propagating SM wave.  The downward propagating SM wave is not excited because we do not allow reflections off our top boundary.  The goal of this paper is to derive how much of the incident wave is reflected as an IGW, and how much is converted into a SM wave.

This can be inferred from figure~\ref{fig:schematic}.  The downward propagating IGW has positive $k_z$.  When it approaches the turning point, it will only interact with the upward propagating SM wave, which also has positive $k_z$.  There cannot be any reflection into an upward propagating IGW wave because that wave has negative $k_z$.  Thus, we expect perfect conversion into SM waves, which is what we find in sections~\ref{sec:constB} \& \ref{sec:variableB}.

This is qualitatively different from typical reflections in the WKB approximation \citep[e.g., what is commonly found in a quantum mechanics course, e.g.,][]{Griffiths95}.  A reflection can occur within the WKB approximation if the wavenumber goes to zero.  In this problem, we have two wavenumbers which approach each other away from zero.  Thus, there cannot be any reflection.

\citet{Cally06} describes the interaction of slow and fast magnetosonic waves.  A slow magnetosonic wave can interact with fast magnetosonic waves in a region in which $c\sim v_A$.  In this case, there are two propagating modes above and below the interaction region.  In the WKB approximation, the local vertical wavenumbers of the waves never equal each other: instead, an avoided crossing takes place.  This is possible because both slow and fast magnetosonic modes have their group velocity parallel to their phase velocity.  Thus, there can be both transmission and conversion between slow and fast modes.

In this section, we have assumed there is a vertical component to the magnetic field.  \citet{macgregor11} studies the horizontal magnetic field problem.  In this singular limit, there are only two wave modes: upward and downward propagating magneto-gravity waves.  This problem can be solved using the WKB approximation.  If the magnetic field becomes sufficiently strong, the vertical wavenumber of the waves go to zero, and the magneto-gravity wave reflects.  This is completely analogous to the WKB reflection problem found in a typical quantum mechanics course.

\section{Variable $N_0$, Constant $\vec{B}_0$}\label{sec:constB}

\begin{figure}\includegraphics[width=\columnwidth]{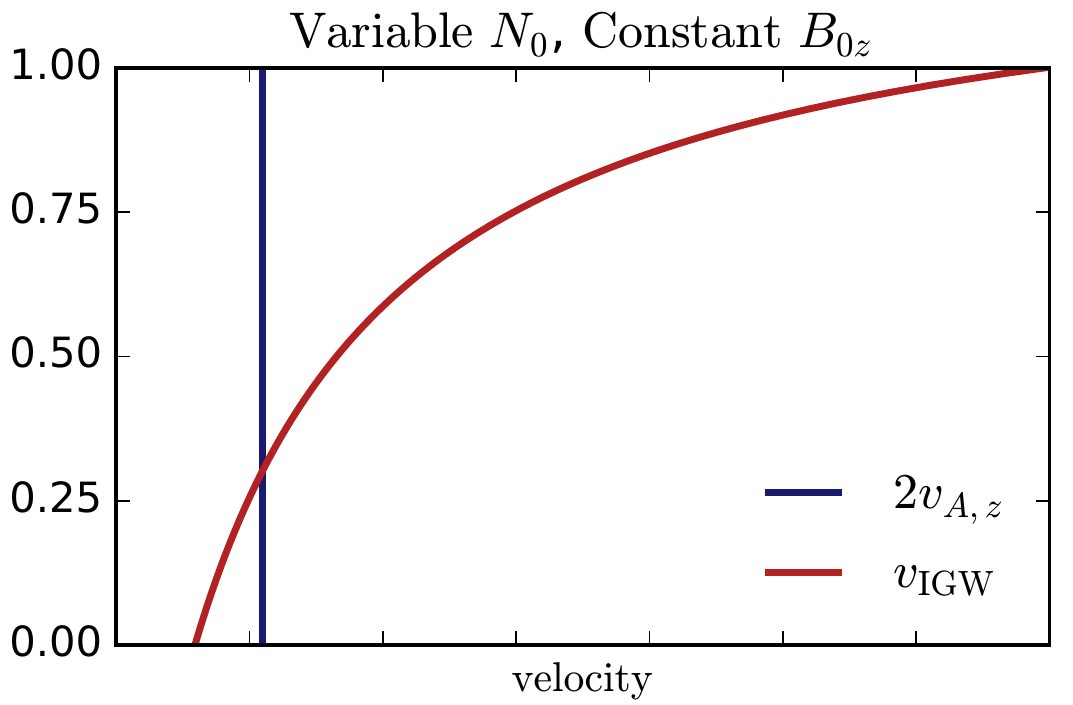}
\caption{\label{fig:background_constB} Profile of the background Alfv\'en velocity and the vertical IGW group velocity for the variable $N_0$, constant $\vec{B}_0$ problem.  The vertical IGW group velocity is inversely proportional to $N_0$.  The waves have a turning point where $2v_{A,z}=v_{\rm IGW}$ (equation~\ref{eqn:constBturning}).}
\end{figure}

We now present specific examples of IGW interaction with magnetic fields. In this section, we solve for the evolution of an IGW propagating downward into a star where $\vec{B}_0$ is constant, but $N_0(z)$ increases with depth.  We assume all wave quantities can be decomposed as
\begin{align}
w(x,y,z,t) = w(z)\exp(i\omega t-i\vec{k}_h\vec{\cdot}\vec{x}),
\end{align}
where $\vec{k}_h$ is the horizontal wavevector.  We parameterize the magnetic field by
\begin{align}
\omega^2_{A,h} & = \frac{(\vec{k}_h\vec{\cdot}\vec{B}_0)^2}{4\pi\rho_0}, \\
v^2_{A,z} & = \frac{B_{0z}^2}{4\pi\rho_0},
\end{align}
which are the Alfv\'en frequency based on the horizontal magnetic field, and the Alfv\'en velocity based on the vertical magnetic field.

In figure~\ref{fig:background_constB} we plot the vertical Alfv\'en velocity along with the vertical IGW group velocity,
\begin{align}
v_{\rm IGW} = \frac{\omega^2}{k_h N_0},
\end{align}
where we have assumed $\omega\ll N_0$.  Our $N_0=2(5-4z)$ increases linearly with depth.  We expect the waves to have a turning point where $v_A\sim v_{\rm IGW}$ (equation~\ref{eqn:heuristic}).

The full solution to this problem is given in appendix~\ref{sec:constBdetails}.  We will include the main results of the calculation here.  We find there is a turning point (see figure~\ref{fig:schematic}) at a height $z_{\rm t}$ satisfying
\begin{align}
2v_{Az}k_hN_0(z_{\rm t}) = v_{Az}^2k_h^2 + \omega^2-\omega_{Ah}^2.
\end{align}
Assuming the local vertical wavenumber is much larger than the horizontal wavenumber, the largest term on the right-hand side is $\omega^2$.  Neglecting the other terms we have
\begin{align}\label{eqn:constBturning}
v_{Az} \approx \frac{1}{2} \frac{\omega^2}{k_h N_0}.
\end{align}
If we substitute $k_h=\sqrt{\ell(\ell+1)}/r$ with $\ell=1$, the critical magnetic field amplitude is
\begin{align}
\label{eqn:constBcritB}
\frac{B_{0z}}{\sqrt{4\pi\rho_0}} \approx \frac{1}{\sqrt{8}} \frac{\omega^2 r}{N_0(z_{\rm t})},
\end{align}
exactly agreeing with the critical magnetic field strength given in \citet{Fuller15}.

Above (and below) the turning point, the equations for the phase and amplitude are given by equations~\ref{eqn:constBphase} \& \ref{eqn:constBamp}.  We plot the local vertical wavenumber as a function of height for the four wave modes in figure~\ref{fig:schematic}.  The WKB amplitude diverges at the turning point, even though the local vertical wavenumber does not go to zero at this point.  This indicates that the WKB solution is not valid near the turning point.

Appendix~\ref{sec:constBdetails} derives the solution near the turning point.  The solution is related to Airy functions, which also appear in more classical WKB turning point problems \citep[e.g.,][]{Griffiths95}.  By asymptotically matching the solution near the turning point to the WKB solution, we find that there is perfect conversion from downward propagating IGWs into upward propagating SM waves, with a $-\pi/2$ phase shift.  This is consistent with the heuristic argument presented in section~\ref{sec:heuristic}, as well as the exact solution for an isothermal atmosphere given in \citet{ZD81}.

\begin{figure}
\includegraphics[width=\columnwidth]{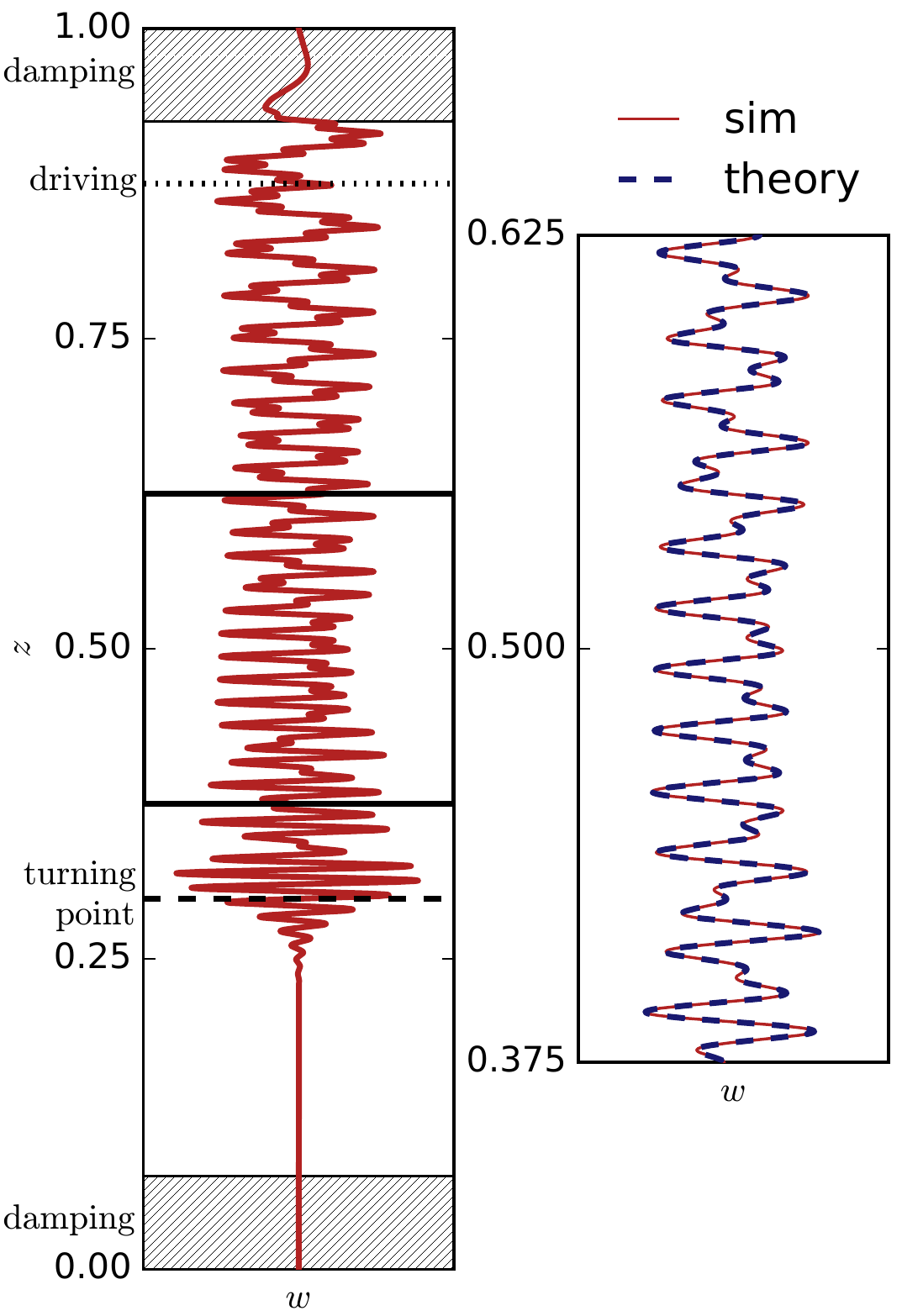}
\caption{\label{fig:constB} Left panel:  A vertical cut of the vertical velocity after the wave simulation (section~\ref{sec:constB}) has reached a steady state.  IGWs are driven at the dotted line, and damped in the dashed regions (given by $z_{\rm top}$ and $z_{\rm bot}$).  The turning point, where the IGWs convert into SM waves, is shown in the dashed line.  Right panel:  A zoom-in of the boxed region from the left panel.  We plot both the simulation field and the theoretical prediction from the WKB approximation.  The excellent agreement confirms that there is perfect conversion from IGWs to SM waves at the turning point with a $-\pi/2$ phase shift.}
\end{figure}

We can test this analytic result by comparing to numerical solutions.  We drive IGWs at the top of the domain as described in section~\ref{sec:setup}.  These waves propagate downward, reach the turning point, convert into SM waves, and then propagate to the top of the domain.  After an initial transient, the system reaches a steady state where the downward IGW flux exactly matches the upward SM wave flux.  A vertical cut of the vertical velocity is shown in the left panel of figure~\ref{fig:constB}.  In the upper part of the domain, there are two dominant oscillation wavelengths, corresponding to the IGW (large wavelength) and the SM wave (short wavelength).  The dashed line shows the turning point---near this point, there is only one dominant oscillation wavelength, as the two waves' vertical wavenumbers approach each other.

We can write the WKB solution as
\begin{align}
w_{\rm WKB} = w_0 \exp(i\phi_0) \left[ A_{\rm IGW} \exp(i\theta_{\rm IGW}) + A_{\rm SM}\exp(i\theta_{\rm SM}+i\Delta\phi)\right],
\end{align}
where $A$ and $\theta$ are the amplitudes and phases of the two waves, given by equations~\ref{eqn:constBphase} \& \ref{eqn:constBamp}.  $w_0$ and $\phi_0$ are the amplitude and overall phase of the solution, and $\Delta\phi$ is the phase difference between the IGW and SM wave, which the theory predicts to be $-\pi/2$.

We perform a nonlinear least squares fit of $w_{\rm WKB}$ to the data between $z=0.67$ and $z=0.76$, solving for $w_0$, $\phi_0$, and $\Delta\phi$.  Although we could have imposed $\Delta\phi=-\pi/2$, we left it as a free parameter to test the theory.  The best fit has $\Delta\phi=-1.5694$, which agrees with $-\pi/2$ to better than $0.1\%$ precision.  We compare the fit to the numerical solution in the right panel of figure~\ref{fig:constB}, and find excellent agreement.  This validates our theoretical prediction that there is complete conversion of IGWs into SM waves with a $-\pi/2$ phase shift.

\section{Variable $\vec{B}_0$, Constant $N_0$}\label{sec:variableB}

\begin{figure}
\includegraphics[width=\columnwidth]{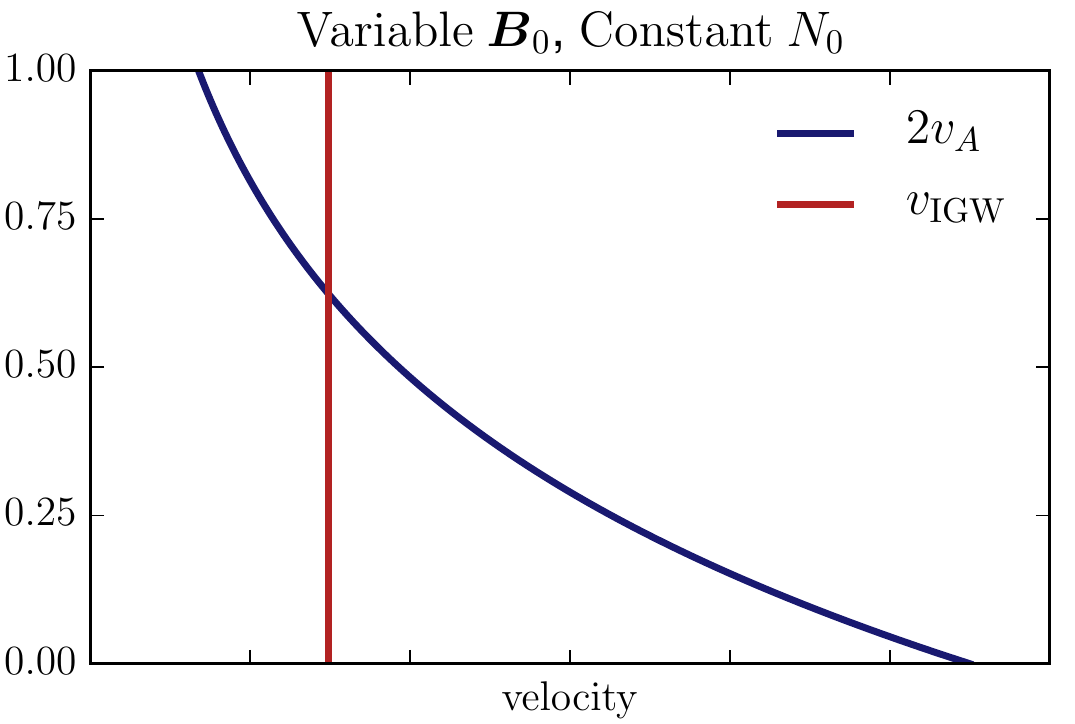}
\caption{\label{fig:background_variableB} Profile of the background Alfv\'en velocity and the vertical IGW group velocity for the variable $\vec{B}_0$, constant $N_0$ problem.  The Alfv\'en velocity (equation~\ref{eqn:alfven_velocity}) decays exponentially with height.  For this problem, there are two turning points (equations~\ref{eqn:variableBturning1} \& \ref{eqn:variableBturning2}), but we only plot twice the Alfv\'en velocity to indicate the approximate positions of the turning points.}
\end{figure}

We now turn our attention to an atmosphere with constant $N_0$, but a magnetic field which is periodic in $x$ and exponentially decaying in $z$ (equation~\ref{eqn:B}).  The background field profiles are plotted in figure~\ref{fig:background_variableB}.  This problem is much more technical, both analytically and numerically.  We summarize the calculation here, but appendix~\ref{sec:variableBdetails} contains the full derivation.

As in section~\ref{sec:constB}, we assume the vertical oscillations are fast, so we can employ the WKB approximation in the vertical direction.  We assume the buoyancy frequency and local vertical wavenumber are both large (order $\epsilon^{-1}$), and that the Alfv\'en velocity is small (order $\epsilon$).  Then to lowest order, the wave equation reduces to equation~\ref{eqn:mathieu}, which for fixed $z$, is equivalent to the Mathieu equation in $x$.  Thus, the solutions are linear combinations of the two Mathieu functions
\begin{align}
M_c^{a,q}(k_B x)\quad {\rm and} \quad M_s^{a,q}(k_B x)
\end{align}
where the parameters $a$ and $q$ depend on $z$ via the local vertical wavenumber and the Alfv\'en velocity (equations~\ref{eqn:a} \& \ref{eqn:q}).  Recall that $k_B$ is the horizontal wavenumber of the background magnetic field.  The solution is not separable as $M_c$ and $M_s$ depend on both $x$ and $z$.

\begin{figure}
\includegraphics[width=\columnwidth]{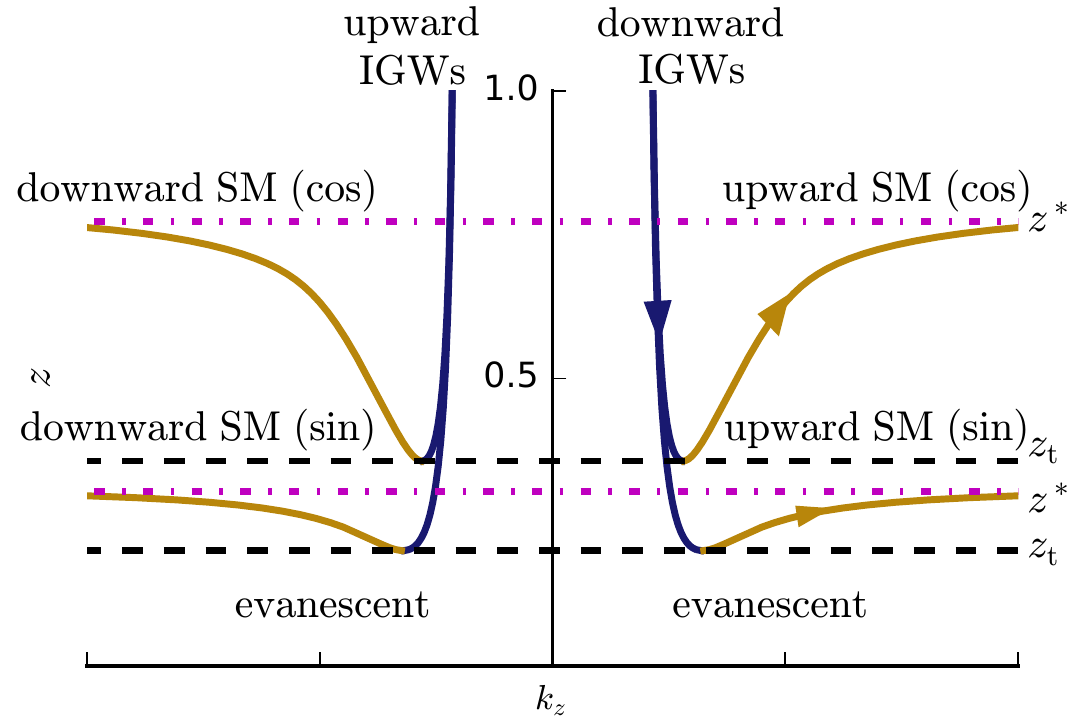}
\caption{\label{fig:kz} The local vertical wavenumber $k_z=\theta'$ in the WKB approximation for the variable $\vec{B}_0$ problem in section~\ref{sec:variableB} with $R=k_h/k_B=2$.  At large $z$, there are four wave modes (blue lines): two IGWs of $\cos$ parity, and two IGWs of $\sin$ parity.  Their vertical wavenumbers are almost identical at large $z$.  There are no propagating waves of a given parity below the turning point $z_{\rm t}$ for that parity (black dashed lines).  When the IGWs approach $z_{\rm t}$, they convert into SM waves of the same parity.  The SM waves' vertical wavenumbers increase with height, and diverge at the Alfv\'en cutoff height $z^*$ (magenta dot-dashed lines).  We expect perfect conversion of downward propagating IGWs into upward propagating SM waves because their wavenumbers equal each other at the turning points.}
\end{figure}

In general, Mathieu functions are not periodic.  They are only periodic for special combinations of $a$ and $q$.  If $a$ equals one of two functions, $A_c(R,q)$ and $A_s(R,q)$, then the appropriate Mathieu function is periodic.  This is the dispersion relation for this problem, and allows us to solve for the local wavenumber at every height.  We plot the local vertical wavenumber, $k_z$, as a function of height for our test problem in figure~\ref{fig:kz}.

For small $q$, corresponding to weak magnetic fields (equation~\ref{eqn:q}),
\begin{align}
M_c^{A_c(R,q),q}(x) \rightarrow \cos(Rk_B x) \, , \\
M_s^{A_s(R,q),q}(x) \rightarrow \sin(Rk_B x) \, ,
\end{align}
where $R=k_h/k_B$ is the ratio of horizontal wavenumbers of the incoming IGW and the background magnetic field.  When IGWs are launched at the top of the domain, they have small $q$ and thus are approximately horizontally sinusoidal with wavenumber $k_h$.  At this stage the vertical wavenumbers of the $\cos$ and $\sin$ parity waves are very similar, so they are nearly indistinguishable in figure~\ref{fig:kz}.  As they propagate downward, $q$ increases, and they develop richer horizontal structure. 

The behavior of the waves depends on their horizontal phase relative to the magnetic field, which is why there is a ``cosine''-like Mathieu functions ($M_c$) and a ``sine''-like Mathieu functions ($M_s$).  In our simulation, we generate traveling waves of both phases, so we excite the two Mathieu functions with equal amplitude.  It is expected that stars would excite both waves to similar amplitudes as well.

\begin{table*}
  \caption{{The non-dimensionalized local vertical wavenumber and Alfv\'en velocity at the turning point $z_{\rm t}$, and the Alfv\'en cutoff velocity, as a function of $R=k_h/k_B$.  For $R$ a non-negative integer, there are different turning points and cutoff points for modes with a $\sin(k_h x)$ parity or a $\cos(k_h x)$ parity.  These are calculated using equation~\ref{eqn:cutoff}, and equations~\ref{eqn:theta_cr} \& \ref{eqn:vA_cr}, together with equation~\ref{eqn:q_cr}.  The values for $R=1/50$ and $R=50$ are representative of the limits $R\rightarrow 0$ and $R\rightarrow\infty$, respectively.}}\label{tab:critical}
{\centering
  \begin{tabular}{ccccccc}
    \hline
    R & $\theta'_{\rm t} \frac{\omega}{k_hN_0}$ ($\cos$ parity) & $v_{A,{\rm t}} \frac{k_hN_0}{\omega^2}$ ($\cos$ parity) & $v_A^* \frac{k_h N_0}{\omega^2}$ ($\cos$ parity) & $\theta'_{\rm t}\frac{\omega}{k_hN_0}$ ($\sin$ parity) & $v_{A,{\rm t}}\frac{k_hN_0}{\omega^2}$ ($\sin$ parity) & $v_A^*\frac{k_h N_0}{\omega^2}$ ($\sin$ parity) \\
    \hline
    \hline
    1/50 & 1.41& 0.707 & 0.020 & 1.41 & 0.707 & 0.020 \\
    1/3 & 1.43 & 0.713 & 0.333 & 1.43 & 0.713 & 0.333 \\
    1/2 & 1.45 & 0.724 & 0.500 & 1.45 & 0.742 & 0.500 \\
    1 & 1.43 & 0.581 & 0.333 & 1.60 & 1.08 & 0.500  \\
    2 & 1.34 & 0.634 & 0.400 & 1.54 & 0.749 & 0.667 \\
    3 & 1.33 & 0.653 & 0.429 & 1.45 & 0.697 & 0.600 \\
    50 & 1.35& 0.671 & 0.495 & 1.35 & 0.671 & 0.505
  \end{tabular}
}
\end{table*}

As the IGW propagates downward, it begins to interact with the magnetic field.  At the turning point $z_{\rm t}$, the IGW and SM wave have equal wavenumbers---this is where mode conversion or reflection can occur.  The critical Alfv\'en velocity $v_{A,{\rm t}}=v_A(z_{\rm t})$ depends on the value of $q$ and $A_p(q,R)$ at the critical height, where the parity $p=c$ or $s$ (equation~\ref{eqn:vA_cr}).  We list several values of $v_{A,{\rm t}}$ for different $R$ in table~\ref{tab:critical}. For a dipole IGW interacting with a dipole magnetic field, $R=1$. The $R=1$ waves of $\cos$ parity have a critical magnetic field strength
\begin{align}
\label{eqn:variableBturning1}
\frac{B_{z,{\rm rms}}}{\sqrt{4\pi\rho_0}} = 0.291 \frac{\omega^2 r}{N_0},
\end{align}
using $k_B=k_h=\sqrt{\ell(\ell+1)}/r$, and where $B_{z,{\rm rms}}$ is the root-mean-square vertical magnetic field at the turning point. The $R=1$ waves with $\sin$ parity have a critical magnetic field strength
\begin{align}
\label{eqn:variableBturning2}
\frac{B_{z,{\rm rms}}}{\sqrt{4\pi\rho_0}} = 0.540 \frac{\omega^2 r}{N_0}.
\end{align}
For field strengths between these two values, only half the IGWs would interact strongly with the magnetic field.  Note that the critical field strengths only vary by a factor of two for any parity and $R$.

At the turning point, the amplitude of the WKB mode diverges.  This indicates that the WKB approximation is no longer valid, because the amplitude is changing too quickly.  Below the turning point, the vertical wavenumber is complex, with both real and imaginary parts, corresponding to evanescent waves.  We derive the solution near the turning point (equation~\ref{eqn:variableBinnersolution}).  As in section~\ref{sec:constB}, we find that the solution is related to Airy functions.  Once again, this means there is perfect conversion from IGWs to SM waves with a $-\pi/2$ phase shift.  This is consistent with the arguments given in section~\ref{sec:heuristic}.  Away from the turning point, we can calculate the amplitude of the wave using equation~\ref{eqn:variableBamplitude}.

As the SM wave propagates upwards, its local vertical wavenumber as predicted by WKB theory increases, since $B_{z,{\rm rms}}$ decreases.  The local vertical wavenumber diverges at a finite Alfv\'en velocity (equation~\ref{eqn:cutoff}, typical values listed in table~\ref{tab:critical}).  This divergence is shown in figure~\ref{fig:kz}.  Taking $k_h=\sqrt{\ell(\ell+1)}/r$ with $\ell=1$, the ``Alfv\'en cutoff velocity'' is
\begin{align}
\label{eqn:cutoff main}
v_A^* = \frac{R}{\sqrt{2}(2R\pm1)} \frac{\omega^2 r}{N_0},
\end{align}
where the plus (minus) sign corresponds to cosine (sine) parity.  Here we assume $R\geq 1$.  If $0<R<1$, the first fraction in equation~\ref{eqn:cutoff main} becomes $R/\sqrt{2}$ (for both parities).  Note that $v_A^*<v_{A,{\rm t}}$, which means the Alfv\'en cutoff point is higher than the turning point.  At the Alfv\'en cutoff velocity, the SM wave's local wavenumber goes to infinity, which indicates that it will probably damp.  This suggests there might be localized wave damping layers at specific radii in the star.  Alternatively, other terms in the evolution which are typically lower order may become important near the Alfv\'en cutoff height, regularizing the problem.

Consider the limits of $R$ very small and $R$ very large.  If $R$ is large, the wave is oscillating rapidly (in the $x$ direction) relative to the magnetic field.  In this limit, one can solve the eigenvalue problem in the $x$ direction (equation~\ref{eqn:mathieu}) using the WKB approximation in $x$.  This is similar to a quantum mechanics problem where the energy is related to the inverse of the local {\it vertical} wavenumber squared, and the potential is due to the sinusoidal magnetic field.  IGWs have small local vertical wavenumbers (i.e., high energy), so they are not affected by the sinusoidal potential.  However, at the turning point, there is global horizontal structure, as the mode has smaller local {\it horizontal} wavenumber near the zeros of the magnetic field, and thus higher amplitude.  Although each local (in $x$) part of the eigenfunction feels an almost constant field, the large-scale field geometry strongly modifies the problem.  Thus, we find that the large $R$ limit does not reduce to the constant magnetic field problem (notice that the cutoff velocity $v_A^*$ is reached at finite height as $R\rightarrow\infty$).

In the limit of $R$ very small, the magnetic field oscillates wildly (in the $x$ direction) relative to the waves.  Even when the wave strongly interacts with the magnetic field at the turning point, the horizontal eigenfunctions are very close to sinusoidal, indicating that the waves effectively feel a constant, averaged magnetic field.  When $R$ is small, $v_A^*$ approaches zero, so there is no Alfv\'en cutoff velocity.  Also, the Alfv\'en velocity at the turning point is equal to $2^{-1/2} \omega^2/(k_hN_0)$, which corresponds to an rms vertical magnetic field of
\begin{align}
\frac{B_{z,{\rm rms}}}{\sqrt{4\pi\rho_0}} = \frac{1}{\sqrt{8}} \frac{\omega^2 r}{N_0},
\end{align}
assuming the IGW is an $\ell=1$ wave.  This is exactly the critical magnetic field strength for the constant magnetic field problem (equation~\ref{eqn:constBcritB}).  In that context, the critical strength is achieved due to variations in $N_0$.  This shows that the small $R$ limit is equivalent to the constant magnetic field case discussed in section~\ref{sec:constB}.

\begin{figure}
\includegraphics[width=\columnwidth]{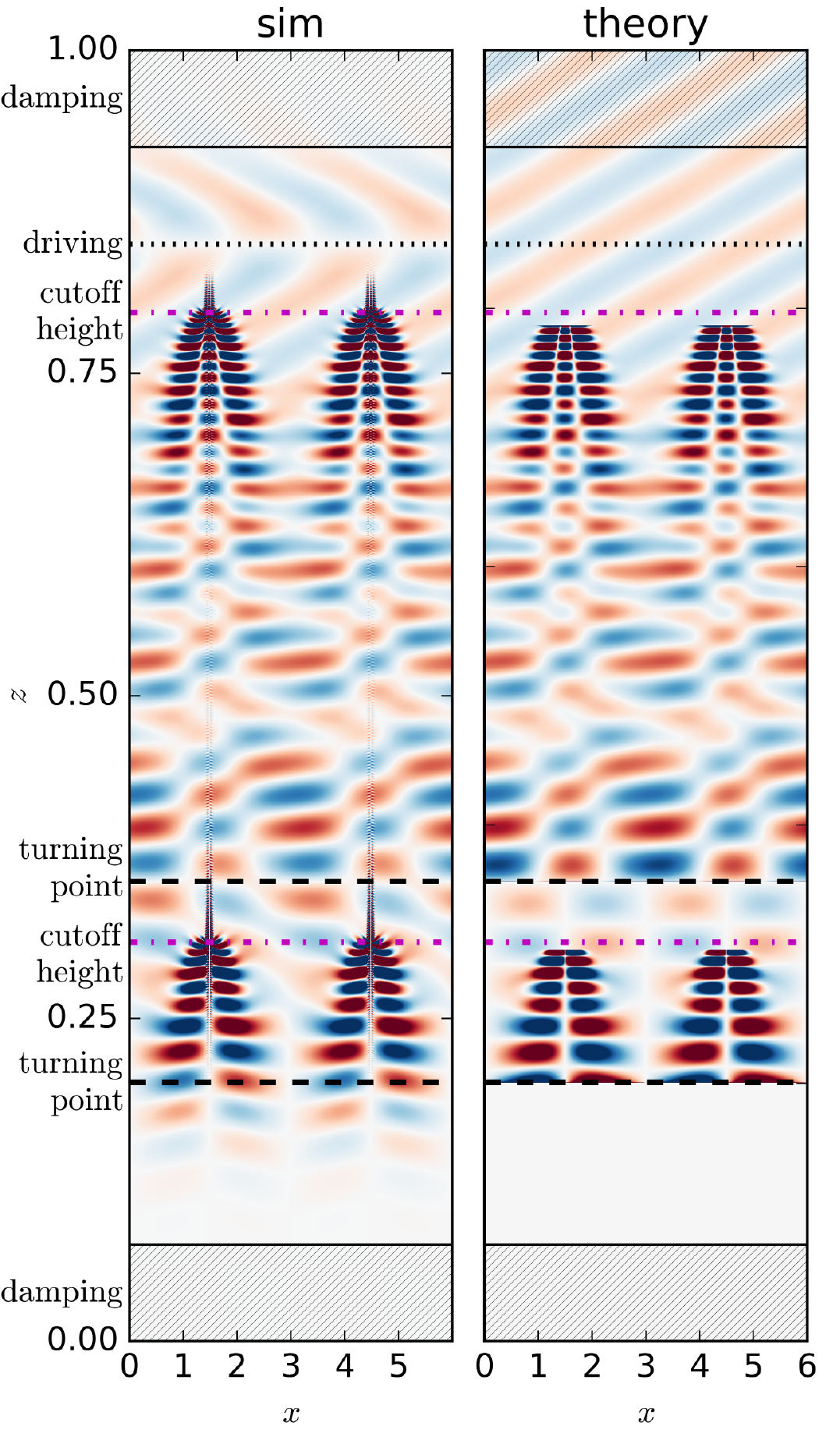}
\caption{\label{fig:variableB} Left panel:  A snapshot of the horizontal velocity after the wave simulation (section~\ref{sec:variableB}) has reached a steady state.  IGWs with $R=k_h/k_B=2$ are driven at the dotted line, and damp in the dashed regions.  The turning point, where the IGWs converts into SM waves, is shown in the thick solid lines.  The analytical theory predicts SM waves reach infinite local wavenumber (and presumably damp) at the Alfv\'en cutoff height, shown in magenta dot-dashed lines.  There are two sets of turning points and Alfv\'en cutoff points, corresponding to modes with either cosine or sine parity.  Right panel:  The horizontal velocity predicted from the WKB approximation.  There is good agreement away from the turning points, driving \& damping layers, and the Alfv\'en cutoff point.  There are high vertical wavenumber fluctuations near $x=1.5$ and $4.5$ in the simulation that are not present in the theory.}
\end{figure}

We now present a simulation of this problem.  Figure~\ref{fig:variableB} (left panel) shows the horizontal velocity, $u$, after the simulation has reached a steady state.  The IGWs are driven at the dotted black line, and have $R=2$.  The IGWs with cosine (sine) parity convert into SM waves at the lower (upper) thick solid black line.  The SM waves with cosine (sine) parity have an Alfv\'en cutoff point at the  lower (upper) magneta dot-dashed line.  This is where the local vertical wavenumber becomes infinite according to our WKB analysis, and we physically expect the waves to damp.

We write the WKB solution as
\begin{align}
u_{\rm WKB} &= u_0 \exp(i\phi_0)\left[F_{c,{\rm IGW}}(z)M_c^{A_c(2,q_{\rm IGW}),q_{\rm IGW}}(k_B x)\exp\left(i\theta_{c,{\rm IGW}}\right)\right. \nonumber \\
&+ F_{c,{\rm SM}}(z)M_c^{A_c(2,q_{\rm SM}),q_{\rm SM}}(k_B x)\exp\left(i\theta_{c,{\rm SM}}+i\Delta\phi_{\rm SM}\right) \nonumber \\
&+ F_{s,{\rm IGW}}(z)M_s^{A_s(2,q_{\rm IGW}),q_{\rm IGW}}(k_B x)\exp\left(i\theta_{s,{\rm IGW}}+i\Delta\phi_{s,{\rm IGW}}\right) \nonumber \\
&+\left. F_{s,{\rm SM}}(z)M_s^{A_s(2,q_{\rm SM}),q_{\rm SM}}(k_B x)\exp\left(i\theta_{s,{\rm SM}}+i\Delta\phi_{\rm SM}\right)\right], \label{eqn:variableBfit}
\end{align}
where $u_0$ and $\phi_0$ are the overall amplitude and phase. The phase of each wave is given by the integral of its local vertical wavenumber, derived from the dispersion relationships~\ref{eqn:Mathieuc} \& \ref{eqn:Mathieus}.  This also determines $q$ by equation~\ref{eqn:q}. The $F$'s are amplitudes given by equation~\ref{eqn:variableBamplitude}.  We only plot the wave modes above the turning points.  Finally, the $\Delta\phi$'s are phase differences relative to the IGW with cosine parity.  The WKB solution predicts $\Delta\phi_{\rm SM}=-\pi/2$ and that $\Delta\phi_{s,{\rm IGW}}=0$.

We perform a nonlinear least squares fit of $u_{\rm WKB}$ to the data at $x=2.25$ and $z$ between $0.51$ and $0.58$.  This determines $u_0$, $\phi_0$, and the two phase differences $\Delta\phi$.  Our fit gives $\Delta\phi_{\rm SM}=-1.48$ and $\Delta\phi_{s,{\rm IGW}}=0.044$, very close to the theoretical prediction of $-\pi/2$ and $0$, respectively.

We can then compare this fit to the data at other $x$ locations and for other heights.  The full 2D field predicted by the theory is shown in the right panel of figure~\ref{fig:variableB}.  Rather than use the Airy function approximation for the solutions near the turning point, we only plot the theoretical velocity above the turning point.  Thus, we do not expect good agreement near the turning points (thick solid lines).  Otherwise, there is good agreement between the theory and the simulation, except near $x=1.5$ and $x=4.5$, which we will discuss below.

\begin{figure}
\includegraphics[width=\columnwidth]{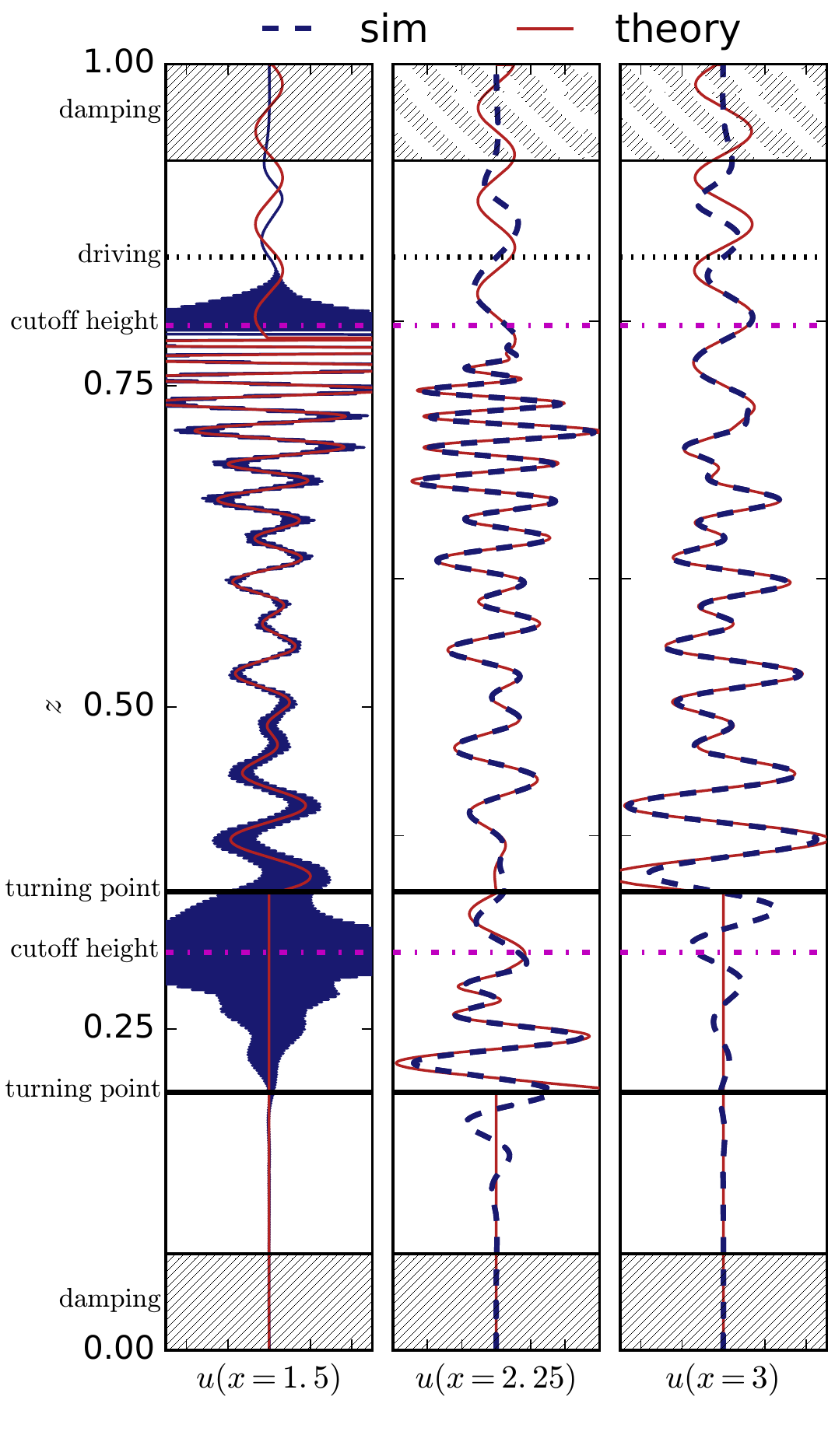}
\caption{\label{fig:variableB_lines} Comparison of the horizontal velocity given by the WKB theory and the simulation, at three $x$ locations.  We fit the WKB solution (equation~\ref{eqn:variableBfit}) to the simulation at $x=2.25$ and between $z=0.51$ and $z=0.58$ to find the overall amplitude, phase, and phase differences between the different wave modes.  The WKB solution is very accurate at other $x$ locations and other heights.  We expect agreement between the turning point \& cutoff height for each parity.  At $x=1.5$, there are small vertical wavelength oscillations not predicted by our theory.  We plot the simulation result at $x=1.5$ in a thin solid line to better visualize the small scale oscillations, which are on lengthscales smaller than the line thickness, leading to an extended opaque region on the plot.}
\end{figure}

Because it is difficult to directly compare the 2D solutions, we plot vertical cuts of the horizontal velocity at three representative $x$ locations in figure~\ref{fig:variableB_lines}.  We show both the simulation result and the theoretical prediction.  We expect the best agreement between the turning point \& Alfv\'en cutoff height for each parity. To improve agreement near the turning points, we could calculate Airy function approximations (equation~\ref{eqn:variableBAiry}), and include the exponentially decaying solutions below the turning points.

However, the largest discrepancy is at $x=1.5$, where there are small vertical lengthscale oscillations in the simulation which are not present in the analytical theory.  These can also be seen in figure~\ref{fig:variableB}.  Near the Alfv\'en cutoff height (magenta dot-dashed line), the waves concentrate near $x=1.5$ and $x=4.5$, where $B_z$ is very close to zero (equation~\ref{eqn:B}).  Near this point, the local vertical wavenumber becomes very large and the analytical solution breaks down.  This may indicate that the asymptotic size of various terms in equation~\ref{eqn:variableBequation} may change near the cutoff height.

Our simulation is well-resolved because we remove power from modes with wavenumber larger than a critical value (this can be viewed as a very aggressive form of hyperdiffusion).  We have repeated the simulation with different critical wavenumbers.  In all cases, the system develops finer and finer scale features until it reaches this critical wavenumber.  We have not found any evidence of an intrinsic lengthscale associated with the Alfv\'en cutoff height.  We hypothesize dissipation may be required to regularize the problem.

Nevertheless, there is good agreement between the simulation and theoretical predictions, as shown in figures~\ref{fig:variableB} \& \ref{fig:variableB_lines}.  Thus, we are confident in our prediction that there is perfect conversion between IGWs and SM waves at the turning points (equations~\ref{eqn:variableBturning1} \& \ref{eqn:variableBturning2}).

\section{Conclusions}\label{sec:conclusions}

In this paper, we study the interaction of internal gravity waves (IGWs) with a magnetic field.  \citet{Fuller15} predicted that interaction will occur when the vertical magnetic field reaches a critical strength (equation~\ref{eqn:heuristic}).  We solve the wave interaction problem using the linear, magneto-Boussinesq equations for two simple cases.  The first has a constant magnetic field, with a buoyancy frequency $N_0^2$ increasing with depth, and the second has a constant $N_0^2$, but a magnetic field which is sinusoidal in the horizontal direction and exponentially decaying with height.  For both problems, we solve the problem analytically using the WKB approximation, and also calculate numerical solutions directly using Dedalus.  We find good agreement between the analytics and the simulations.

In section~\ref{sec:heuristic}, we argue that an IGW propagating into a region of strong magnetic field (or large buoyancy frequency) should completely convert into a slow magnetosonic (SM) wave, i.e., an Alfv\'enic wave.  Wave interaction occurs when two waves have similar frequency and wavevector.  Crucially, a downward propagating IGW has a {\it positive} vertical wavenumber.  Thus, this IGW can convert into an upward propagating SM wave, which also has a positive vertical wavenumber.  There is no reflection into upward propagating IGWs, which have {\it negative} vertical wavenumbers.

In both magnetic configurations we investigate, there is perfect conversion from IGWs to SM waves at a turning point.  Using the WKB approximation, and properties of Mathieu functions (for the variable magnetic field case), we find a turning point at a critical magnetic field strength.  At this turning point, the solution is well approximated by Airy functions, as in the normal WKB theory near a turning point.  However, this is qualitatively different from the normal WKB turning point because the wavenumber never approaches zero.  Because of the Airy function behavior, we find perfect conversion from IGWs to SM waves with a $-\pi/2$ phase shift in both problems.  We find excellent agreement between the analytic and numerical solutions.

Our work has important implications for the asteroseismic signatures of stars with strong internal magnetic fields. In red giant stars, \citet{Fuller15} suggested that IGWs interacting with a strong core magnetic field would be scattered into either high multipole IGWs or SM waves. Our results show that the latter process dominates, and IGWs interacting with a strong field will generally be converted into SM waves which will likely dissipate upon traveling into unmagnetized regions of the star (i.e., where the Alfv\'en velocity is smaller than the cutoff value given in equation~\ref{eqn:cutoff main}). Thus, the magnetic greenhouse effect discussed by \citet{Fuller15} is not as important as the conversion into SM waves and their subsequent dissipation. However, the observational signature is the same: IGWs penetrating into magnetized red giant cores are likely to be totally damped. The amplitudes of dipole modes in red giant stars with magnetic cores should therefore be suppressed as predicted by \citet{Fuller15}, and we expect these stars to exhibit only envelope acoustic modes (no mixed modes) in their pulsation spectra. Stars whose pulsation spectra clearly show mixed modes (even with low amplitudes) are unlikely to have strong core magnetic fields, and they require an alternative mode suppression mechanism.

Another important consequence of our results is that any strongly magnetized star will not exhibit g mode pulsations, because IGW will be converted into damped SM waves. Our specific examples give a critical magnetic field strength consistent (up to a factor of two) with \citet{Cantiello_2016}. This could explain why there are very few magnetic white dwarfs that are known g mode pulsators.\footnote{The few magnetic pulsating white dwarfs of which we are aware (e.g., \cite{Dufour_2008}) are rare DQ white dwarfs with carbon-dominated atmospheres. A preliminary model indicates a magnetic field of $\sim 10^7 \, {\rm G}$ would be sufficient to suppress g mode pulsations of frequencies $f=1 \, {\rm mHz}$ in a hydrogen atmosphere ZZ-Ceti star, but a stronger field would be required to suppress g modes in DQ white dwarfs which lack composition gradients near their surface. The measured field from \cite{Dufour_2008} is only $\sim 10^6 \ {\rm G}$ and may not be strong enough to suppress the g mode pulsation at $\sim 2 \, {\rm mHz}$ in that star.}
Additionally, g mode pulsations in $\gamma$-Doradus, sdB, or SPB stars can be totally suppressed by strong internal magnetic fields, and non-pulsating stars within these respective instability strips are good candidates for harboring internal fields. 

Future work should extend these results to three dimensional spherical geometry.  The two dimensional cartesian geometry used in this paper makes the problem simpler, but is only applicable for waves with horizontal and vertical lengthscales much smaller than their local propagation radius in their star, $r$.  However, dipole IGWs have horizontal wavelengths comparable to $r$, and the problem is generally three dimensional (due to three directions determined by gravity, the magnetic field, and the wave vector). It is possible that some three-dimensional field configurations will allow for slightly different dynamics, but we expect our main results should still hold in three dimensions and in spherical geometry via the argument made in section~\ref{sec:heuristic}.

\section*{Acknowledgments}

\noindent{}The authors would like to thank Lars Bildsten, Eliot Quataert, Ellen Zweibel, Anna Lieb, Stephane Mathis, Dennis Stello, Rafael Garcia, and Frank Timmes for helpful discussions.  D.L. is supported by the Hertz Foundation, a PCTS fellowship, and a Lyman Spitzer Jr.~fellowship, and would like to thank the University of Sydney School of Mathematics and Statistics for helping fund a visit to Sydney.  This work has been carried out in the framework of the Labex MEC (ANR-10-LABX-0092) and of the A*MIDEX project (ANR-11-IDEX-0001-02), funded by the ``Investissements d'Avenir'' French Government program managed by the French National Research Agency (ANR).  G.M.V. acknowledges support from the Australian Research Council, project number DE140101960.  This research is funded in part by the Gordon and Betty Moore Foundation through Grant GBMF5076 to Lars Bildsten, Eliot Quataert and E. Sterl Phinney.  This research was supported in part by the National Science Foundation under Grant No. NSF PHY-1125915.  We thank KITP for supporting a follow-up meeting where much of this work was initiated.  Resources supporting this work were provided by the NASA High-End Computing (HEC) Program through the NASA Advanced Supercomputing (NAS) Division at Ames Research Center.  This project was supported by NASA under the SPIDER TCAN, grant number NNX14AB53G.

\appendix

\section{Analytic Solution to Variable $N_0$, Constant $\vec{B}_0$ Problem}\label{sec:constBdetails}

The dispersion relation for the problem described in section~\ref{sec:constB} is
\begin{align}
&\left[v_{Az}^2\partial_z^4 + \left(\omega^2-\omega_{Ah}^2 - v_{Az}^2k_h^2\right)\partial_z^2\right. \nonumber \\
&\quad\quad\quad+ \left. \left(N_0^2(z)+\omega_{Ah}^2 - \omega^2\right)k_\perp^2\right] w(z) = 0.\label{eqn:constBDR}
\end{align}
This is quadratic in $\partial_z^2$, and the turning point $z_{\rm t}$ occurs where the discriminant is zero,
\begin{align}
2v_{Az}k_hN_0(z_{\rm t}) = v_{Az}^2k_h^2 + \omega^2-\omega_{Ah}^2.
\end{align}
Here we assume $v_{Az}\neq 0$.  The horizontal magnetic field case \citep{macgregor11} is singular as there are only two vertical modes if $v_{Az}=0$; otherwise, there are four.

We can non-dimensionalize the problem by defining
\begin{align}
\zeta&=(z-z_{\rm t})k_h \sqrt{\frac{N_0}{k_h v_{Az}} -1}, \\
C(\zeta)&=\frac{N_0^2(z_{\rm t})-N_0^2(\zeta)}{(N_0(z_{\rm t})-v_{Az}k_h)^2},
\end{align}
so that equation~\ref{eqn:constBDR} becomes
\begin{align}
(\partial_\zeta^2+1)^2 w(\zeta) - C(\zeta) w(\zeta) = 0.
\end{align}

To apply the WKB approximation, we assume $C$ varies on a large lengthscale $Z\equiv \zeta\epsilon$.  The wave oscillation is on the short lengthscale $\zeta$.  In terms of $Z$, the equation becomes
\begin{align}
\left(\epsilon^2\partial_Z^2 + 1\right)^2 w(Z) - C(Z)w(Z)=0.
\end{align}
Now we can apply the WKB ansatz
\begin{align}
w(Z)=A(Z)\exp\left(\frac{i\theta(Z)}{\epsilon}\right).
\end{align}
The equation for the phase is
\begin{align}
\label{eqn:constBphase}
(\theta'^2-1)^2+C = 0,
\end{align}
where a prime denotes derivative with respect to $\zeta$.  The solution to the next-order equation for the amplitude is
\begin{align}
\label{eqn:constBamp}
A=\frac{1}{\sqrt{\theta'(\theta'^2-1)}}.
\end{align}
The amplitude divergences at $z_{\rm t}$, where $\theta'^2$ approaches unity.  This indicates the presence of an inner solution.  The inner solution determines the fate of the downward propagating IGW when it approaches the turning point.

Near $Z=0$, we can expand $C=SZ$.  We define
\begin{align}
w(Z)=W(Z)\exp(\pm i Z/\epsilon),
\end{align}
so the dispersion relation becomes
\begin{align}
\label{eqn:constBinnerDR}
\epsilon^4\partial_Z^4W\pm4i\epsilon^3\partial_Z^3W-4\epsilon^2\partial_Z^2W-SZ W =0.
\end{align}
On the inner length scale $\eta=Z/\epsilon^{2/3}$, the last two terms balance to leading order, and we are left with
\begin{align}
\partial_\eta^2 W = -\frac{S}{4}\eta W,
\end{align}
which is the Airy equation.

The full solution in the inner region is an Airy function multiplied by $\exp(\pm i Z/\epsilon)$.  We're interested in the first Airy function which will decay to zero below the turning point.  We have
\begin{align}
w&(Z)\sim {\rm Ai}\left(-Z \left(\frac{\sqrt{S}}{2\epsilon}\right)^{2/3}\right) e^{\pm i Z/\epsilon} \sim \sin\left(\frac{\pi}{4}+\frac{\sqrt{S}}{3}\frac{Z^{3/2}}{\epsilon}\right)e^{\pm i Z/\epsilon} \nonumber \\
&\sim \exp\left(i \frac{\pm Z+\sqrt{S}Z^{3/2}/3}{\epsilon}+\frac{i\pi}{4}\right) - \exp\left(i \frac{\pm Z-\sqrt{S}Z^{3/2}/3}{\epsilon}-\frac{i\pi}{4}\right).
\end{align}
Taking the plus sign, one can check that the first exponential term matches asymptotically to the upward propagating SM wave, and the second exponential term matches asymptotically to the downward propagating IGW.  The amplitudes of the two terms are equal, so there is perfect conversion from IGW to SM wave.  The SM wave has a phase shift of $-\pi/2$ compared to the IGW.

Equation~\ref{eqn:constBinnerDR} can be solved directly via the method of steepest descent.  We recover the same solution: perfect conversion with a $-\pi/2$ phase shift.  In this case, the phase shift is due to the $90$ degree angle between the steepest descent curves through the critical points (in the phase variable) corresponding to the two wave modes above the turning point.

\section{Analytic Solution to Variable $\vec{B}_0$, Constant $N_0$ Problem}\label{sec:variableBdetails}

First, we write several quantities in terms of potentials,
\begin{align}
\vec{u} & = \vec{\nabla}\vec{\times}(\vec{e}_y \psi), \\
\vec{B} & = \vec{\nabla}\vec{\times}(\vec{e}_y \varphi), \\
\vec{B}_0 &= \vec{\nabla}\vec{\times}(\vec{e}_y \varphi_0),
\end{align}
such that the magneto-Boussinesq equations become
\begin{align}
\rho & = -\frac{\rho_0N_0^2}{i\omega g}J(z,\psi), \\
\varphi & = -\frac{J(\psi,\varphi_0)}{i\omega}, \\
i\rho_0\omega\nabla^2\psi & = J(\varphi_0,\nabla^2 \varphi) + g J(z,\rho),
\end{align}
where $J(f,g)\equiv\partial_xf \partial_z g - \partial_xg \partial_z f$.  For our definition of $\vec{B}_0$ (equation~\ref{eqn:B}), the magnetic potential is
\begin{align}
\varphi_0 = \frac{B_0}{k_B}\sin(k_B x)\exp(-k_B z)
\end{align}
These can be combined to derive the equation
\begin{align}
&\omega^2\partial_z^2\psi + (\omega^2-N_0^2)\partial_x^2\psi + v_A^2\cos(k_B x)^2\partial_z^4\psi \nonumber \\
&+v_A^2\left[2k_B^2\partial_z^2\psi - k_B(2+\cos(2 k_B x))\partial_z^3\psi + 3k_B\cos(2k_B x)\partial_x^2\partial_z\psi \nonumber \right. \\
& - 2 k_B^2\partial_x^2\psi - 2k_B \partial_x^2\partial_z\psi + \partial_x^2\partial_z^2\psi + \sin(k_B x)^2\partial_x^4\psi \nonumber \\
&\left. + \sin(2k_B x)(-3k_B\partial_x\partial_z^2\psi + \partial_x\partial_z^3\psi + k_B\partial_x^3\psi + \partial_x^3\partial_z\psi)\right] = 0, \label{eqn:variableBequation}
\end{align}
where
\begin{align}\label{eqn:alfven_velocity}
v_A^2 = \frac{B_0^2}{4\pi\rho_0} \exp(-2k_B z).
\end{align}

We now assume $N^2\sim \epsilon^{-2}$ is large and $v_A^2\sim \epsilon^2$ is small.  If $\omega$ and $k_h$ are order unity, then equations~\ref{eqn:IGWdisp} \& \ref{eqn:SMdisp} are asymptotically consistent if the vertical wavenumber is large, order $\epsilon^{-1}$.  Thus, we search for a solution of the form
\begin{align}
\psi = M(x,z) e^{i\theta(z)/\epsilon},
\end{align}
For notational simplicity, we use primes to denote $z$ derivatives of $\theta$, i.e., $\theta'=\partial_z\theta$.  Then to lowest order, equation~\ref{eqn:variableBequation} becomes
\begin{align}
\label{eqn:mathieu}
-\omega^2\theta'^2 M - N_0^2\partial_x^2M + v_A^2\cos(k_B x)^2\theta'^4 M = 0,
\end{align}
which is an eigenvalue equation in $x$ for $M$.  The solution is
\begin{align}
M(x,z) = F_c(z) M_c^{a,q}(k_B x) + F_s(z)M_s^{a,q}(k_B x),
\end{align}
where the $M_p^{a,q}$ are Mathieu functions with parity $p=c$ for cosine or $s$ for sine, and normalization given in equation~\ref{eqn:mathieunormalization}.
The $F$'s are amplitude functions to be determined at next-to-leading order, and
\begin{align}
\label{eqn:a}
a &= \frac{\theta'^2(\omega^2-\theta'^2v_A^2/2)}{k_B^2N_0^2}, \\
q &= \frac{\theta'^4v_A^2}{4k_B^2N_0^2}.\label{eqn:q}
\end{align}
Below we use various properties of Mathieu functions and their characteristic values which can be found in, e.g., \citet{Olver10}.

In general, Mathieu functions are not periodic.  However, we require solutions which are periodic in $x$.  The Mathieu functions are only periodic if $a$ is equal to a Mathieu characteristic value,
\begin{align}
\label{eqn:Mathieuc}
\frac{\theta'^2(\omega^2-\theta'^2v_A^2/2)}{k_B^2N_0^2}=A_c\left(R,\frac{\theta'^4v_A^2}{4k_B^2N_0^2}\right), \\
\frac{\theta'^2(\omega^2-\theta'^2v_A^2/2)}{k_B^2N_0^2}=A_s\left(R,\frac{\theta'^4v_A^2}{4k_B^2N_0^2}\right), \label{eqn:Mathieus}
\end{align}
where $A_c(R,q)$ is the characteristic value for $M_c^{a,q}(k_B x)$, and $A_s(R,q)$ is the characteristic value for $M_s^{a,q}(k_B x)$.  In the limit $q\rightarrow 0$, $M_c^{A_c(R,q),q}(k_B x)$ approaches $\cos(R k_B x)$, and $M_s^{A_s(R,q),q}(k_B x)$ approaches $\sin(R k_B x)$.  Thus, $R$ represents the ratio of the IGW's wavenumber to the magnetic field's wavenumber, $R=k_h/k_B$.  The functions $A_p(R,q)$ are discontinuous at integer values of $R$, but are equal to each other at non-integer values.  If $R$ is an integer, $A_s(R,q)$ is the limit from the left as $R$ approaches the integer, and $A_c(R,q)$ is the limit from the right.

Equations~\ref{eqn:Mathieuc} \& \ref{eqn:Mathieus} are the dispersion relation for this problem,
\begin{align}
\label{eqn:variableBdisp}
D_p\left(z,\theta'^2\right) = \frac{v_A^2}{2}\theta'^4 - \omega^2\theta'^2+k_B^2N_0^2 A_p\left(R;z,\theta'^2\right) = 0.
\end{align}
Consider equation~\ref{eqn:variableBdisp} as an equation in $\theta'^2$ for every height $z$.  Then, for a given height we can count the number of roots $\theta'^2$.  At $\theta'^2=0$, we have that $A_p(R,q)= R^2$.  Thus,
\begin{align}
D_p(z,0) = k_B^2 N_0^2 R^2 > 0.
\end{align}
As $\theta'\rightarrow\infty$, we have
\begin{align}\label{eqn:Dc1}
D_c\left(z,\theta'^2\right) \approx \left[(2R+1)k_BN_0v_A-\omega^2\right]\theta'^2 - \frac{1}{4}(2R^2+2R+1), \\
D_s\left(z,\theta'^2\right) \approx \left[(2R-1)k_BN_0v_A-\omega^2\right]\theta'^2 - \frac{1}{4}(2R^2-2R+1),
\end{align}
for $R$ a positive integer.  If $R^{-1}$ is an integer between zero and one, then both $D_p$ are given by equation~\ref{eqn:Dc1} setting $R=0$.  We define the Alfv\'en cutoff velocity when $R$ is a positive integer by
\begin{align}\label{eqn:cutoff}
v_{A\pm}^* = \frac{1}{2R\pm1} \frac{\omega^2}{k_BN_0}.
\end{align}
If $R^{-1}$ is an integer between zero and one, then the first fraction of equation~\ref{eqn:cutoff} is equal to one.  If $v_A>v_{A\pm}^*$, then $D_p$ becomes positive at large $\theta'^2$.  Since it is also positive at $\theta'^2=0$, there must be an even number of roots of $\theta'^2$.  In practice, we find either two (double) roots (two IGWs and two SM waves), or zero roots (only evanescent modes).  If $v_A<v_{A\pm}^*$, then $D_p$ becomes negative at large $\theta'^2$, which means there are an odd number of roots.  In practice, we find only one (double) root, corresponding to two IGWs.  Thus, there are no SM waves above the height corresponding to the Alfv\'en cutoff velocity.

Next, we derive the turning point $z_{\rm t}$ at which the IGWs convert to SM waves.  This occurs at the Alfv\'en velocity $v_{A,{\rm t}}$ at which the dispersion relation has a double root, i.e., $D_p(z_{\rm t},\theta'^2)=0$ and $\partial_{\theta'^2}D_p(z_{\rm t},\theta'^2)=0$.  The latter condition is
\begin{align}
\label{eqn:doubleroot}
v_{A,{\rm t}}^2 \theta'^2\left[2+\partial_qA_p\left(R,\frac{\theta'^4 v_{A,{\rm t}}^2}{4 k_B^2N_0^2}\right)\right] = 2 \omega^2.
\end{align}
Using that $D_p$ is also zero, we find the double root is at $q_{\rm t}$ given by
\begin{align}
\label{eqn:q_cr}
A_p\left(R,q_{\rm t}\right) = 2 q_{\rm t} \left[1+\partial_q A_p\left(R,q_{\rm t}\right)\right].
\end{align}
For $R$ a non-negative integer, $q_{\rm t}$ takes different values for $p=c$ and $p=s$.  However, we suppress the dependence on $p$ for notational simplicity.

We are guaranteed to have at least one solution to this equation.  When $q$ is small, the left-hand side approaches $R^2$, whereas the right-hand side approaches $0$, so the left-hand side is greater.  For $q$ large and $R$ an integer,
\begin{align}
&A_p(R,q)\rightarrow -2q + 2(2R\pm1)\sqrt{q} - \frac{1}{8}\left[(2R\pm1)^2+1\right] +\ldots, \\
&2q\left[1+\partial_qA_p(R,q)\right]\rightarrow -2q + 2(2R\pm1)\sqrt{q}+\mathcal{O}\left(q^{-1/2}\right),
\end{align}
where $p=c$ corresponds to the plus sign, and $p=s$ corresponds to the minus sign.  If $R^{-1}$ is an integer between zero and one, replace $(2R\pm1)$ with one.  Thus, the left-hand side is smaller than the right-hand side, and there must be at least one root $q_{\rm t}$.  When we have $q_{\rm t}$, the values of $\theta'^2$ and $v_A^2$ are
\begin{align}
\label{eqn:theta_cr}
\theta'^2_{\rm t} & = \frac{2 k_B^2 N_0^2}{\omega^2}q_{\rm t}\left[2+\partial_qA_p\left(R,q_{\rm t}\right)\right], \\
v^2_{A,{\rm t}} &= \frac{\omega^4}{k_B^2 N_0^2 q_{\rm t}\left[2+\partial_qA_p\left(R,q_{\rm t}\right)\right]^2}.\label{eqn:vA_cr}
\end{align}
We calculate these for several values of $R$ in table~\ref{tab:critical}.

In figure~\ref{fig:kz}, we plot $k_z=\theta'$ as a function of height for the example problem discussed in section~\ref{fig:variableB}.  At large heights (where $q$ is small), the roots of the dispersion relation are insensitive to parity.  Because we use $R=2$, there are two turning points for modes with either $\cos$ or $\sin$ parity.  At the turning points, IGWs convert to SM waves.  The local vertical wavenumbers of the SM waves increase with height until they diverge at $z^*$.

Up until this point, we have only used the properties of the Mathieu characteristic values.  Before deriving the behavior near the turning point, we derive useful identities about the Mathieu functions themselves.  Define the inner product
\begin{align}
\left\langle f,g\right\rangle_R = \int_0^{L_R} f g dx,
\end{align}
where $L_R=2\pi\times\max(1,1/R)$, assuming $R$ or $1/R$ is a non-negative integer.  The Mathieu functions are orthonormal with normalization of one half,
\begin{align}\label{eqn:mathieunormalization}
\left\langle M_{p_1}^{A_{p_1}(R_1,q),q}(x),M_{p_2}^{A_{p_2}(R_2,q),q}(x)\right\rangle_{R_1} = \frac{1}{2}\delta_{p_1,p_2}\delta_{R_1,R_2},
\end{align}
where $\delta$ is the Kronecker delta.

The Mathieu functions satisfy
\begin{align}
\partial_x^2 M_p^{A_p(R,q),q}(x) + \left[A_p(R,q)-2q\cos(2 x)\right]M_p^{A_p(R,q),q}(x) = 0.
\end{align}
Taking a derivative with respect to $q$, we find
\begin{align}
&\left[\partial_x^2+A_p(R,q)-2q\cos(2x)\right]\partial_qM_p^{A_p(R,q),q}(x) \nonumber \\
&+ \left[\partial_q A_p(R,q) - 2 \cos(2x)\right]M_p^{A_p(R,q),q}(x) = 0.
\end{align}
Now we take the inner product of this equation with $M_p^{A_p(R,q),q}(x)$.  There is no contribution from the top row because the Mathieu operator is self-adjoint.  Thus, we find
\begin{align}
\left\langle M_p^{A_p(R,q),q},\cos(2x)M_p^{A_p(R,q),q}\right\rangle_R = \frac{1}{4}\partial_q A_p(R,q).
\end{align}
This implies
\begin{align}
\left\langle M_p^{A_p(R,q),q},\partial_x^2M_p^{A_p(R,q),q}\right\rangle_R = -\frac{A_p(R,q)}{2}+\frac{q\partial_q A_p(R,q)}{2}.
\end{align}

We now calculate the behavior of the waves near the turning point.  Below we will show that the WKB amplitude diverges at the turning point.  This indicates the need for an inner solution.  For simplicity, we only consider the cosine parity mode.  Near the turning point, the Alfv\'en velocity is
\begin{align}
v_A^2 \approx v^2_{A,{\rm t}} + \delta v_A^2 = v^2_{A,{\rm t}} + z \partial_z v_A^2 = v^2_{A,\rm t}\left[1 + \frac{d\log v_A^2}{dz} \epsilon^{2/3} \eta\right],
\end{align}
where $\eta=z/\epsilon^{2/3}$ is the inner lengthscale on which the amplitude varies.  Our ansatz for the solution to equation~\ref{eqn:variableBequation} is
\begin{align}\label{eqn:variableBinnersolution}
\psi(x,z) = \Upsilon_c\left(\frac{z}{\epsilon^{2/3}}\right) M_c^{A_c(R,q_{\rm t}),q_{\rm t}}(k_B x) e^{i\theta_{\rm t}(z)/\epsilon}.
\end{align}
The amplitude function $\Upsilon_c$ evolves on the inner lengthscale $\eta=z/\epsilon^{2/3}$.  We can substitute this into equation~\ref{eqn:variableBequation}, recalling that $N_0$ is order $\epsilon^{-1}$ and $v_A$ is order $\epsilon$.  Next project onto $M_c^{A_c(R,q_{\rm t}),q_{\rm t}}(k_B x)$.

The resulting equation is satisfied to lowest order ($\epsilon^{-2}$) because it is the Mathieu equation, and the lowest order solution is a Mathieu function.  The next-to-leading order ($\epsilon^{-5/3}$) equation is
\begin{align}
\left[\omega^2-v_{A,{\rm t}}^2\theta'^2_{\rm t} - \frac{v_{A,{\rm t}}^2\theta'^2_{\rm t}}{2} \partial_q A_c(R,q)|_{q_{\rm t}}\right] 2i\theta'_{\rm t}\partial_\eta\Upsilon_c=0,
\end{align}
which is automatically satisfied by equation~\ref{eqn:doubleroot}.

The next order ($\epsilon^{-4/3}$) terms give us the equation for $\Upsilon_c(\eta)$,
\begin{align}
\label{eqn:variableBAiry}
\partial_\eta^2\Upsilon_c = -\frac{k_B\theta'^2_{\rm t}}{2} \eta \Upsilon_c,
\end{align}
using $d\log v_A^2/dz=-2k_B$.  This is the Airy equation, so $\Upsilon_c$ must be the sum of Airy functions.  The argument given in appendix~\ref{sec:constBdetails} can be applied here to show that there is perfect conversion from IGWs into SM waves with a $-\pi/2$ phase shift.

All that remains is to derive the amplitude equation.  We can expand $\psi$ as
\begin{align}
\psi = \sum_{p} \psi_0(R,p;x,z)e^{i\theta_p^{(R)}(z)/\epsilon} + \epsilon \psi_1(x,z) e^{i\theta_1(z)/\epsilon},
\end{align}
where
\begin{align}
\psi_0(R,p;x,z) = F_p^{(R)}(z) M_p^{A_p(R,q),q}(x)
\end{align}
and $\psi_1(x,z)$ is the first order correction.  To derive the amplitude equation, we substitute this into equation~\ref{eqn:variableBequation}, take the order $\epsilon^{-1}$ terms, and then project out with $M_p^{A_p(r,q),q}(x)$.  To prevent the first order correction from resonating with the zeroth order solution, we require
\begin{align}
&\left\langle M_p^{A_p(R,q),q}(x),\left[ \omega^2\left( \psi_0 \theta''+2\theta'\partial_z\psi_0\right)\right.\right. \nonumber \\
&-v_A^2\cos(k_Bx)^2\left(4\partial_z\psi_0\theta'^3+6\psi_0\theta''\theta'\right) \nonumber \\
&\left.\left.+v_A^2 \theta'^3 \left(k_B (2+\cos(2k_Bx))\psi_0 - \sin(2k_Bx)\partial_x\psi_0\right) \right] \right\rangle_R=0,
\end{align}
where we have dropped $R$ and $p$ labels for brevity.  Using the inner product identities, as well as integration by parts, we find
\begin{align}
\label{eqn:hardamplitude}
&4\theta'\left[\omega^2-v_A^2\theta'^2\left(1+\frac{\partial_qA_p(R,q)}{2}\right)\right]\partial_z \log F_p^{(R)}(z) \nonumber \\
&= -\left[ 2\left(\omega^2-3v_A^2\theta'^2\right)\theta''+4v_A^2k_B\theta'^3 \right. \nonumber \\
&\left. v_A^2\left(2k_B\theta'-3\theta''\right)\theta'^2 \partial_qA_p(R,q) - v_A^2 \theta'^3 \partial_z \partial_qA_p(R,q)\right].
\end{align}
The amplitude $F_p^{(R)}(z)$ diverges when the bracketed term on the first line is equal to zero.  But this is exactly the double root condition (equation~\ref{eqn:doubleroot}), so the amplitude diverges at the turning point.

To evaluate the amplitude more generally, it is useful to derive an expression for $\partial_qA_p(R,q)$ by taking a $z$ derivative of the dispersion relation (equations~\ref{eqn:Mathieuc} \& \ref{eqn:Mathieus})
\begin{align}
\partial_qA_p(R,q) = \frac{4\theta''(\omega^2-\theta'^2v_A^2)-2\theta'^3v_A'v_A}{v_A\theta'^2(2\theta''v_A+\theta'v_A')}.
\end{align}
Substituting this into equation~\ref{eqn:hardamplitude}, and using that $v_A\sim \exp(-k_Bz)$, we find a very simple relation for the amplitude,
\begin{align}
\label{eqn:variableBamplitude}
\partial_z\left[\frac{F_p^{(R)}\theta'}{\sqrt{k_B\theta'-2\theta''}}\right]=0.
\end{align}

\section{Equation Implementation in Dedalus}\label{sec:dedalusequations}

Our implementation of equations~\ref{eqn:momentum}--\ref{eqn:induction} in Dedalus is different for the two problems described in this paper.  For the problem with variable $N_0$, but constant $\vec{B}_0$ (section~\ref{sec:constB}), we use
\begin{align}
\partial_t \rho' - N_0^2 w & = F - \rho D_N, \\
\partial_x u+\partial_z w & = 0, \\
\partial_t u + \partial_x p + (\partial_x B_z - \partial_z B_x) B_0 & = - u D_N, \\
\partial_t w + \partial_z p + \rho' & = -w D_N, \\
\partial_t B_z + B_0 \partial_x u & =  - B_z D_N, \label{eqn:Bz}\\
\partial_x B_x + B_{zz} & = 0, \label{eqn:divB}\\
\partial_z B_z - B_{zz} & = 0, \label{eqn:Bzz}
\end{align}
where $\rho'=g\rho/\rho_0$ is the normalized density perturbation.  The constant magnetic field is assumed to be in the vertical direction, and has magnitude $B_0$.  The magnetic field equations are singular for the horizontally averaged mode---for this mode, we replace equations~\ref{eqn:Bz}--\ref{eqn:Bzz} with
\begin{align}
\partial_t B_x - B_0 u_z & = - B_x D_N, \\
B_z & = 0, \\
\partial_z u - u_z & = 0.
\end{align}
The boundary conditions are $w=B_x=0$ on the top boundary, and $w=B_z=0$ on the bottom boundary.  For the horizontally averaged mode, our boundary conditions are $p=B_x=0$ on the top boundary, and $w=u=0$ on the bottom boundary.

For the problem with variable $\vec{B}_0$ but constant $N_0$ (section~\ref{sec:variableB}), we use
\begin{align}
\partial_t \rho' - N_0^2 w & = F - \rho D_N, \\
\partial_x u+\partial_z w & = 0, \\
\partial_t u + \partial_x p & =  (\partial_z B_x - \partial_x B_z) B_{0z}- u D_N, \\
\partial_t w + \partial_z p + \rho' & = -(\partial_z B_x - \partial_x B_z) B_{0x} -w D_N, \\
\partial_t B_z & = B_{0x} \partial_x w - B_{0z} \partial_x u - u \partial_x B_{0z} - w\partial_z B_{0z} - B_z D_N, \label{eqn:Bz2}\\
\partial_x B_x + \partial_z B_z & = 0, \label{eqn:divB2}
\end{align}
where $B_{0x}$ and $B_{0z}$ are the $x$ and $z$ components of $\vec{B}_0$.  We replace the horizontally \& vertically averaged mode of the divergence equation with the condition that the domain-averaged pressure is zero.  As above, the magnetic field equations are singular for the horizontally averaged mode---for this mode, we replace equations~\ref{eqn:Bz2} \& \ref{eqn:divB2} with
\begin{align}
B_z & = 0, \\
\partial_t B_x & = B_{0x} \partial_x u + B_{0z}\partial_z u - u\partial_x B_{0x} - w \partial_z B_{0x} - Bx D_N.
\end{align}
Because the simulation is periodic in the $z$ direction, we do not need to impose boundary conditions.  For both implementations, all terms on the left-hand (right-hand) side of the equals sign are treated implicitly (explicitly) in our timestepping scheme.

\bibliographystyle{mn2e}
\bibliography{converted_to_latex}

\label{lastpage}
\end{document}